\documentclass[fleqn,usenatbib]{mnras}

\pdfoutput=1

\usepackage{amssymb,amsmath,latexsym,mathrsfs}
\usepackage{graphicx}
\usepackage{epsfig}
\usepackage{varioref}
\usepackage{hyperref}
\usepackage{color}
\usepackage{multirow}
\usepackage{array}
\usepackage{wasysym}
\usepackage{color}
\usepackage{float}
\usepackage{mathtools}
\usepackage[utf8]{inputenc}
\usepackage[T1]{fontenc}
\usepackage{footnote}

\begin{document}

\title[CCH to calibrate the ladders and measure $\Omega_k$]{Cosmic chronometers to calibrate the ladders and measure the curvature of the Universe. A model-independent study}
\author[A. Favale, A. G\'omez-Valent \& M. Migliaccio]
{Arianna Favale\thanks{arianna.favale@roma2.infn.it}, Adri\`a G\'omez-Valent\thanks{agvalent@roma2.infn.it} and Marina Migliaccio\thanks{migliaccio@roma2.infn.it}
 \\
Dipartimento di Fisica, Università di Roma Tor Vergata, via della Ricerca Scientifica, 1, 00133, Roma, Italy\\
INFN, Sezione di Roma 2, Università di Roma Tor Vergata, via della Ricerca Scientifica, 1, 00133 Roma, Italy
}

\date{\today}

\maketitle

\begin{abstract}
We use the state-of-the-art data on cosmic chronometers (CCH) and the Pantheon+ compilation of supernovae of Type Ia (SNIa) to test the constancy of the SNIa absolute magnitude, $M$, and the robustness of the cosmological principle (CP) at $z\lesssim 2$ with a model-agnostic approach. We do so by reconstructing $M(z)$ and the curvature parameter $\Omega_{k}(z)$ using Gaussian Processes. Moreover, we use CCH in combination with data on baryon acoustic oscillations (BAO) from various galaxy surveys (6dFGS, BOSS, eBOSS, WiggleZ, DES Y3) to measure the sound horizon at the baryon-drag epoch, $r_d$, from each BAO data point and check their consistency. Given the precision allowed by the CCH, we find that $M(z)$, $\Omega_k(z)$ and $r_d(z)$ are fully compatible (at $<68\%$ C.L.) with constant values. This justifies our final analyses, in which we put constraints on these constant parameters under the validity of the CP, the metric description of gravity and standard physics in the vicinity of the stellar objects, but otherwise in a model-independent way. If we exclude the SNIa contained in the host galaxies employed by SH0ES, our results read $M=(-19.314^{+0.086}_{-0.108})$ mag, $r_d=(142.3\pm 5.3)$ Mpc and  $\Omega_k=-0.07^{+0.12}_{-0.15}$, with $H_0=(71.5\pm 3.1)$ km/s/Mpc (68\% C.L.). These values are independent from the main data sets involved in the $H_0$ tension, namely, the cosmic microwave background and the first two rungs of the cosmic distance ladder. If, instead, we also consider the SNIa in the host galaxies, calibrated with Cepheids, we measure $M=(-19.252^{+0.024}_{-0.036})$ mag, $r_d=(141.9^{+5.6}_{-4.9})$ Mpc, $\Omega_k=-0.10^{+0.12}_{-0.15}$ and $H_0=(74.0^{+0.9}_{-1.0})$ km/s/Mpc.
 
\end{abstract}



\section{Introduction}\label{sec:intro}

The absolute distance and time scales in cosmology are set by the Hubble-Lema\^itre constant, $H_0$, which also sets the energy scale of the universe's expansion through the Friedmann equation. Its accurate determination is therefore of utmost importance and has been a long-pursued goal since the very birth of modern (relativistic) cosmology and the idea of an expanding universe, almost one century ago \citep{Hubble:1929}. Yet, we still do not have a consensus value for this parameter. 

The SH0ES collaboration has measured $H_0$ making use of the distance ladder method. They employ 42 supernovae of Type Ia (SNIa) contained in host galaxies with Cepheids at $z<0.01$, i.e. at distances $d\lesssim 40$ Mpc, to calibrate the absolute magnitude of SNIa, $M^{R22}=(-19.253\pm 0.027)$ mag. By extending the ladder to the Hubble flow, up to $z\sim  0.15$, i.e. $d\sim 600$ Mpc, they obtain $H^{R22}_0=(73.04\pm 1.04)$ km/s/Mpc \citep{Riess:2021jrx}. The distance ladder measurement is basically model-independent, since it only relies on the Cosmological Principle (CP) and the assumption that SNIa are good enough standardizable objects, i.e., with a standardized $M$ which remains constant from our vicinity to the far end of the Hubble flow. 

Cosmic microwave background (CMB) observations, on the other hand, allow us to measure in a model-independent way and very precisely the position of the first acoustic peak of the CMB temperature angular power spectrum or, equivalently, the angle $\theta_*=r_*/D_M(z_*)$, where $r_*$ is the comoving sound horizon at recombination and $D_M(z_*)=(1+z_*)D_A(z_*)$ the comoving angular diameter distance to the last-scattering surface. However, these two quantities, $r_*$ and $D_M(z_*)$, cannot be obtained separately with a model-agnostic method. Pre-recombination physics, which depends of course on the model, fixes $r_*$ and this, in turn, fixes $D_M(z_*)$ to fulfill the tight constraint on $\theta_*$\footnote{The {\it Planck} collaboration has measured the CMB acoustic angular scale to $0.03\%$ precision, $100\theta_*=1.04110\pm 0.00031$ \citep{Aghanim:2018eyx}.}. 
In the context of the $\Lambda$CDM, the fit to the full TT,TE,EE+lensing CMB likelihood from {\it Planck} leads to $H^{P18}_0=(67.36\pm 0.54)$ km/s/Mpc \citep{Aghanim:2018eyx}. The latter is in $\sim 5\sigma$ tension with SH0ES. This constitutes the well-known $H_0$ tension, the biggest mismatch between the standard model of cosmology and current observations, see \citep{Verde:2019ivm,Perivolaropoulos:2021jda,Abdalla:2022yfr} for dedicated reviews. 

The angle $\theta_*$ is the CMB analogue of the transverse baryon acoustic oscillations (BAO) scale, $r_d/D_M(z)$, which has been measured by several galaxy surveys, with $r_d$ the comoving sound horizon at the baryon-drag epoch and $z$ being in this case the characteristic redshift of the survey. As $r_*$, $r_d$ is also set by the physics in the pre-recombination era. Considering on top of CMB, data from BAO and uncalibrated low- and high-redshift SNIa one gets $r_d = (147.17\pm 0.20)$ Mpc and $M=(-19.403\pm 0.010)$ mag in the standard cosmological model \citep{Gomez-Valent:2022hkb}. This value of $M$ is again in $\sim 5\sigma$ tension with the one reported by SH0ES, as expected, since their large determination of $H_0$ is induced by the large value of $M$ ($M^{R22}$) obtained from the calibration in the host galaxies. A prior on $r_d$ derived from CMB analyses or a precise estimate of the primordial deuterium abundance can be used to calibrate the BAO distances and, consequently, also other low-redshift observables like SNIa, by assuming standard physics before decoupling. This, in turn, can be employed to extract a model-dependent estimate of $H_0$ and is the basis of the so-called inverse distance ladder, which leads again to a small value of the Hubble parameter, very close to the {\it Planck}/$\Lambda$CDM value \citep{Aubourg:2014yra,Cuesta:2014asa,Addison:2017fdm,DES:2017txv,Feeney:2018mkj}. We remark that this method only allows for a model-dependent determination of $H_0$, even when no specific cosmological model is assumed at late times by using, for example, cosmography.

In view of the above discussion, it is clear that the $H_0$ tension can be recast in a tension in the calibrators of the direct and inverse distance ladders, $M$ and $r_d$. These quantities play a crucial role in the Hubble tension (see e.g. \citealt{Bernal:2016gxb,Aylor:2018drw,Camarena:2019moy,Camarena:2019rmj}). It is therefore very important to measure these distance calibrators independently from the CMB and the first rungs of the direct distance ladder, as a means of cross-checking the results obtained with the standard methods described above. 

Apart from that, it is also interesting to perform these calibrations in a model-independent way. Many models have been proposed in the last years to alleviate the $H_0$ tension: coupled dark energy models \citep{Pettorino:2013oxa,Gomez-Valent:2020mqn,Agrawal:2019dlm,Archidiacono:2022iuu,Goh:2022gxo}, modified gravity \citep{SolaPeracaula:2019zsl,SolaPeracaula:2020vpg,Ballesteros:2020sik,Braglia:2020iik,Braglia:2020auw,Benevento:2022cql}, running vacuum models \citep{SolaPeracaula:2021gxi}, early dark energy \citep{Poulin:2018cxd,Niedermann:2019olb,Agrawal:2019lmo,Hill:2020osr,Gomez-Valent:2021cbe,Gomez-Valent:2022bku}, scenarios with varying atomic constants \citep{Liu:2019awo,Sekiguchi:2020teg,Lee:2022gzh}, or models with primordial magnetic fields \citep{Jedamzik:2020krr}. See \citep{DiValentino:2021izs} for a review and a more complete list of references. The vast majority of these proposals introduce some kind of new physics in the last stages of the recombination epoch, triggering shifts in the value of $r_d$ accompanied also by changes at low redshift to keep the good description of the CMB and BAO data. Other authors have suggested an ultra-late time transition in the effective gravitational coupling and hence in $M$ at $z_t\approx 0.01$ to loosen the tension, \citep{Marra:2021fvf,Perivolaropoulos:2022vql,Perivolaropoulos:2022khd}. We could use the model-independent estimation of the distance calibrators to assess the viability of these models beyond $\Lambda$CDM. Thus, it is clear that calibrating the ladders using independent methods and following model-independent approaches can be very relevant. The results obtained with these alternative methods could be employed to shed some light into the discussion, potentially arbitrating the Hubble tension itself.

In this paper we use the state-of-the-art data on cosmic chronometers (CCH) to calibrate the cosmic ladders and measure the curvature of the universe in a model-independent framework, employing also the Pantheon$+$ compilation of SNIa and BAO data from various galaxy surveys (6dFGS, BOSS, eBOSS, WiggleZ, DES Y3). The original idea of this calibration technique was presented in \citep{Sutherland:2012ys}. It was applied for the first time by \cite{Heavens:2014rja} and subsequently employed in several works in the light of new data and different statistical methods, see e.g. \citep{Verde:2016ccp,Haridasu:2018gqm,Dhawan:2021mel,Gomez-Valent:2021hda}. It assumes that gravity can be described by a metric theory, together with the CP and the validity of CCH as reliable cosmic clocks, and SNIa and BAO as optimal standard candles and standard rulers, respectively. Here, we reconstruct the shape of $H(z)$ from CCH and the one of the apparent magnitude of SNIa $m(z)$ with Gaussian Processes (GPs) and use them to test some of these very basic assumptions, which are usually taken for granted in other works. In particular, we reconstruct $\Omega_k(z)$ applying the method proposed by \cite{Clarkson:2007pz} to test the homogeneity property of the universe by checking that this function is compatible with a constant for $z\lesssim2$. See \citep{Cai:2015pia,Yu:2016gmd,Liu:2020pfa} for similar studies along this direction. We also reconstruct the absolute magnitude of SNIa as a function of the redshift, $M(z)$, and check that no evolution is preferred by current data. This analysis is on the lines of the one by \cite{Benisty:2022psx}, but we use different data sets, have a better control of the effect of correlations and get rid of double-counting issues. Finally, we perform a consistency test among the BAO data points employed in this paper, and show that according to the low-redshift data sets under consideration, there is no significant statistical tension between them. 

All in all, these preliminary tests legitimize the final part of this work, in which we obtain model-independent constraints on $\Omega_k$ and the calibrators $M$ and $r_d$, which are also independent of the main drivers of the Hubble tension. This independent calibration of the ladders is obviously relevant for the discussion of the $H_0$ tension for the reasons already explained. $\Omega_k$, on the other hand, provides us with information about the early universe and the period of inflation. It is a pivotal parameter. In the context of $\Lambda$CDM, the CMB data from {\it Planck} prefer a closed universe at $\gtrsim 2\sigma$ C.L. for the {\it Planck} TT,TE,EE likelihood, $\Omega_k=-0.044^{+0.018}_{-0.015}$ ($68\%$ C.L.), and at a slightly lower level when also the CMB lensing information is included in the analysis, $\Omega_k=-0.0106\pm 0.0065$ \citep{Aghanim:2018eyx,Handley:2019tkm,DiValentino:2019qzk}. However, when data on BAO, SNIa, the full-shape galaxy power spectrum or CCH are added on top of CMB, this deviation from spatial flatness disappears \citep{Aghanim:2018eyx,Efstathiou:2020wem,Vagnozzi:2020rcz,Vagnozzi:2020dfn}. Same conclusions are reached when CMB data from the Atacama Cosmology Telescope are employed alone or in combination with WMAP \citep{ACT:2020gnv}. For a review, we refer the reader to \citep{DiValentino:2020srs}. See also the exhaustive work by \cite{deCruzPerez:2022hfr} for constraints on the curvature in non-flat $\Lambda$CDM and its extensions under a large variety of data sets, and \citep{Collett:2019hrr} for a cosmographical measurement of $H_0$ and $\Omega_k$ from SNIa and strong lensing data. In this paper we measure the curvature parameter without assuming any cosmological model.

This manuscript is organized as follows. In Sec. \ref{sec:data} we describe in detail the low-$z$ data sets employed throughout the paper, namely CCH, SNIa and BAO. In Sec. \ref{sec:GP} we remind the reader what a Gaussian Process is and explain some of its novel and useful technical aspects, e.g. on how to select a kernel applying an objective mathematical criterion. We reconstruct the shape of $H(z)$ and $m(z)$, which is important for the subsequent parts of the paper. In Sec. \ref{sec:tests} we perform the preliminary tests already mentioned in the previous paragraphs, and in Sec. \ref{sec:CalibSec} we calibrate the ladders and measure the curvature of the universe using different data set combinations. We also discuss how our constraints improve if we decrease the uncertainties of the CCH data.  In Sec. \ref{sec:conclusions} we finally provide our conclusions. 


\begin{table}
    \centering
    \caption{List with the 32 CCH data points on $H(z)$ used in this work, obtained from the references quoted in the last column. In the case of Refs. \citep{moresco2012improved, Moresco:2016mzx}, the central values of $H(z)$ are computed by performing the arithmetic mean of the measurements obtained with the BC03 \citep{Bruzual:2003tq} and MaStro \citep{Maraston:2011sq} SPS models. The covariance matrix is computed using the method presented in \citep{Moresco:2020fbm}, which incorporates both the statistical and systematic errors. See the main text and the aforesaid references for details.}
    \begin{tabular}{lccr}
        \hline
        $z$ & $H(z)$ [Km/s/Mpc] & References \\
        \hline
        \hline
        0.07 & 69.0$\pm$19.6 & \cite{Zhang:2012mp} \\ 
        0.09 & 69.0$\pm$12.0 & \cite{Jimenez:2003iv} \\
        0.12 & 68.6$\pm$26.2 & \cite{Zhang:2012mp} \\
        0.17 & 83.0$\pm$8.0 & \cite{Simon:2004tf} \\
        0.1791 & 78.0$\pm$6.2 & \cite{moresco2012improved} \\
        0.1993 & 78.0$\pm$6.9 & \cite{moresco2012improved} \\
        0.2 & 72.9$\pm$29.6 & \cite{Zhang:2012mp} \\
        0.27 & 77.0$\pm$14.0 & \cite{Simon:2004tf}  \\
        0.28 & 88.8$\pm$36.6 &\cite{Zhang:2012mp} \\
        0.3519 & 85.5$\pm$15.7 & \cite{moresco2012improved} \\
        0.3802 & 86.2$\pm$14.6 & \cite{Moresco:2016mzx} \\
        0.4 & 95.0$\pm$17.0 & \cite{Simon:2004tf}  \\
        0.4004 & 79.9$\pm$11.4 & \cite{Moresco:2016mzx} \\
        0.4247 & 90.4$\pm$12.8 & \cite{Moresco:2016mzx} \\
        0.4497 & 96.3$\pm$14.4 & \cite{Moresco:2016mzx} \\
        0.47 & 89.0$\pm$49.6 & \cite{Ratsimbazafy:2017vga} \\
        0.4783 & 83.8$\pm$10.2 & \cite{Moresco:2016mzx} \\
        0.48 & 97.0$\pm$62.0 & \cite{Stern:2009ep} \\
        0.5929 & 107.0$\pm$15.5 & \cite{moresco2012improved} \\
        0.6797 & 95.0$\pm$10.5 & \cite{moresco2012improved} \\
        0.75 & 98.8$\pm$33.6 & \cite{Borghi:2021rft} \\
        0.7812 & 96.5$\pm$12.5 & \cite{moresco2012improved} \\
        0.8754 & 124.5$\pm$17.4 & \cite{moresco2012improved} \\
        0.88 & 90.0$\pm$40.0 & \cite{Stern:2009ep} \\
        0.9 & 117.0$\pm$23.0 & \cite{Simon:2004tf}  \\
        1.037 & 133.5$\pm$17.6 & \cite{moresco2012improved} \\
        1.3 & 168.0$\pm$17.0 & \cite{Simon:2004tf}  \\
        1.363 & 160.0$\pm$33.8 & \cite{Moresco:2015cya} \\
        1.43 & 177.0$\pm$18.0 & \cite{Simon:2004tf}  \\
        1.53 & 140.0$\pm$14.0 & \cite{Simon:2004tf}  \\
        1.75 & 202.0$\pm$40.0 & \cite{Simon:2004tf}  \\
        1.965 & 186.5$\pm$50.6 & \cite{Moresco:2015cya} \\
        \hline
    \end{tabular}\label{tab:CCH}
\end{table}

\section{Data} \label{sec:data}

We dedicate this section to describe the low-redshift data sets employed in this study.


\subsection{Cosmic chronometers}\label{sec:dataCCH}
Massive passively evolving galaxies with old stellar populations and very low star formation rates, i.e. with very little contamination from young components, can be employed as cosmic chronometers using the so-called differential age technique. The idea dates back to the seminal work by \cite{Jimenez:2001gg} and is based on the fact that in a Friedmann-Lema\^itre-Robertson-Walker (FLRW) universe the Hubble function can be written as

\begin{equation}\label{eq:H(z)}
    H(z)=-\frac{1}{1+z}\frac{dz}{dt}\,,
\end{equation}
with $dt/dz$ the look-back time differential change with redshift. Passively evolving galaxies formed at high redshift ($z\sim2-3$) and over a very short period of time ($t\sim0.3$ Gyr). By comparing two ensembles of galaxies that formed at the same time but with different (close enough) redshifts, it is possible to estimate the derivative $dz/dt$ using their spectra and a stellar population synthesis (SPS) model. This, in turn, allows us to measure $H(z)$, under the assumption that General Relativity and standard physics hold in the environment of the stars. Apart from that and the CP\footnote{The expression \eqref{eq:H(z)} might hold even in the presence of cosmic backreaction  \citep{Koksbang:2021qqc}. Cosmic distances, though, would depart from the FLRW ones, so our analyses of Secs. \ref{sec:testM}, \ref{sec:testBAO}, and \ref{sec:CalibSec} are strictly valid under the assumption of the CP, i.e. if the impact of the backreaction is negligible. See the aforesaid sections for details.}, the CCH data are free from other cosmological assumptions, what makes these data very suitable to perform model-independent analyses like those we will carry out in this work. In addition, direct measurements of $H(z)$ can be employed to calibrate the ladders, since they set the energy scale in the universe. In our study, CCH will play an analogous role to the calibrated Cepheids employed by SH0ES in the direct distance ladder.

We provide the list with the 32 CCH data points employed in this paper in Table \ref{tab:CCH}, together with the original references. They span over the redshift range $0.07<z<1.965$ and constitute the most updated data set on CCH in the literature. In the last years important efforts have been dedicated to build the error budget of the CCH data, see e.g. \citep{Moresco:2020fbm}. The full (non-diagonal) covariance matrix of the data is computed as \footnote{\url{https://gitlab.com/mmoresco/CCcovariance}}: 

\begin{equation}\label{eq:covM}
C_{ij}=C_{ij}^{stat}+C_{ij}^{sys}\,.
\end{equation}
$C^{stat}$ contains the statistical errors and is diagonal. The systematic uncertainties contained in $C^{sys}$ account for several effects related to the estimate of physical properties of the galaxies, e.g. the stellar metallicity and the possible contamination by a young component, which are uncorrelated for objects at different redshifts. This is not the case for other sources of uncertainty, as they are primarily due to the choice of initial mass function, stellar library, etc., which rely on the common SPS model used to study the evolution of galaxies. See again \citep{Moresco:2020fbm} for a more detailed account of the origin and modeling of systematic errors in the CCH data.


\subsection{Supernovae of Type Ia}\label{sec:dataSNIa}

We make use of the Pantheon$+$ compilation of Type Ia supernovae \citep{Scolnic:2021amr}, which includes 1701 light curves of 1550 unique, spectroscopically confirmed SNIa, ranging in redshift from $z = 0.001$ to $2.26$ and coming from 18 different surveys \footnote{\url{https://github.com/PantheonPlusSH0ES/DataRelease}}. The main changes with respect to the original Pantheon compilation from \citep{Pan-STARRS1:2017jku} are that in Pantheon$+$ the sample size (especially at $z<0.01$) and the redshift span are larger, and there has also been an improved treatment of systematic uncertainties in redshifts, peculiar
velocities, photometric calibration, and intrinsic-scatter models of SNIa. In particular, we would like to remark that due to some cuts, not all the SNIa contained in Pantheon are found in the improved Pantheon$+$ compilation. There are some redshift ranges in which the number of SNIa is smaller, cf. Fig. 1 of \citep{Scolnic:2021amr}.

In this paper we actually use two different SNIa samples. In our main analyses we remove the data points from the SNIa that are contained in the host galaxies of SH0ES \citep{Riess:2021jrx,Brout:2022vxf} in order to obtain results independent of them. The remaining sample contains 1624 data points. In Sec. \ref{sec:SNIahost} we also use the full Pantheon$+$ compilation together with the distances to the host galaxies obtained by SH0ES in the first rungs of the distance ladder to assess their impact in our model-independent measurement of $M$, $r_d$ and $\Omega_k$.

The SNIa data are given as follows. For each lightcurve we have the apparent magnitude as measured on Earth, $\tilde{m}$, together with the heliocentric and Hubble diagram redshifts, denoted as $z_{\rm hel}$ and $z_{\rm HD}$ \citep{Carr:2021lcj}, respectively. If $M$ is the standardized absolute magnitude of the SNIa and $\tilde{D}_L$ is the luminosity distance inferred from the measurements for a fixed $M$, we have the following relation, 

\begin{equation}
\tilde{m}(z_{\rm hel},z_{\rm HD}) = M +25+5\log_{10}\left(\frac{\tilde{D}_L(z_{\rm hel},z_{\rm HD})}{1\,{\rm Mpc}}\right)\,,
\end{equation}
with 

\begin{equation}
\tilde{D}_L(z_{\rm hel},z_{\rm HD})=\left(\frac{1+z_{\rm hel}}{1+z_{\rm HD}}\right) D_L(z_{\rm HD})\,,
\end{equation}
and

\begin{equation}\label{eq:DL}
D_L(z) = \frac{c(1+z)}{\sqrt{\Omega_k H_0^2}}\sinh\left(\sqrt{\Omega_k H_0^2}\int_0^z\frac{dz^\prime}{H(z^\prime)}\right)\,,
\end{equation}
where $\Omega_k=-kc^2/(R_0H_0)^2$ is the curvature density parameter, with $k=0,-1,+1$ for a flat, open and closed universe, respectively. $R_0$ is a constant with units of length that can be interpreted as the current radius of curvature in a closed universe. Using these relations, it is possible to rewrite the expression of the apparent magnitude in the most usual form, only in terms of the redshift $z_{\rm HD}$,   

\begin{align}\label{eq:mfinal}
m(z_{\rm HD}) &=  \tilde{m}(z_{\rm hel},z_{\rm HD})-5\log_{10}\left(\frac{1+z_{\rm hel}}{1+z_{\rm HD}}\right)\nonumber\\
&=M +25+5\log_{10}\left(\frac{D_L(z_{\rm HD})}{1\,{\rm Mpc}}\right)\,.
\end{align}
This is the apparent magnitude that would be measured in absence of peculiar motions, and is the function we will reconstruct in Sec. \ref{sec:GP} to perform the tests of Secs. \ref{sec:testM} and \ref{sec:testCP}. We consider in all our analyses the effect of statistical and systematic uncertainties in the Pantheon+ data through the corresponding non-diagonal covariance matrix.


\begin{table*}
    \centering
    \caption{List with the 11 BAO data points used in this work. The fiducial values of the comoving sound horizon appearing in the third column are $r_{d}^{fid} = 147.5$ Mpc for \citep{Carter_2018} and $r_{d}^{fid} = 148.6$ Mpc for \citep{Kazin_2014}. We have duly taken into account the existing correlations between the data points of WiggleZ, BOSS DR12 and eBOSS DR16. See the quoted references and the text in Sec. \ref{sec:dataBAO} for details.}
    \label{tab:BAO_data}
    \begin{tabular}{ccccc}
        \hline
        Survey & $z$ & Observable & Measurement & References \\
        \hline
        \hline
        6dFGS+SDSS MGS & 0.122 & $D_V(r_{d}^{fid}/r_d)$  & $539\pm17$ [Mpc] & \cite{Carter_2018} \\
        \hline
         WiggleZ & 0.44 & $D_V(r_{d}^{fid}/r_d)$  & $1716.4\pm83.1$ [Mpc] & \cite{Kazin_2014}\\
         & 0.60 & $D_V(r_{d}^{fid}/r_d)$  & $2220.8\pm100.6$ [Mpc] & \\
         & 0.73 & $D_V(r_{d}^{fid}/r_d)$  & $2516.1\pm86.1$ [Mpc] & \\
         \hline
        BOSS DR12 & 0.32 & $r_{d}H/(10^{3} km/s)$ & $11.549\pm0.385$ & \cite{Gil_Mar_n_2016}\\
         & & $D_A/r_d$ & $6.5986\pm0.1337$ &\\
         & 0.57 & $r_{d}H/(10^{3} km/s)$ & $14.021\pm0.225$ & \\
         & & $D_A/r_d$ & $9.389\pm0.103$ &\\
         \hline
         DES Y3 & 0.835 & $D_M/r_d$ & $18.92\pm0.51$ & \cite{Abbott_2022}\\
         \hline
         eBOSS DR16 & 1.48 & $D_M/r_d$ & $30.21 \pm 0.79$ & \cite{Neveux_2020}\\
         
         &  & $c/r_{d}H$ & $13.23 \pm 0.47$ & \cite{Hou_2020}
        \\
        \hline
    \end{tabular}
\end{table*}

\subsection{Baryon Acoustic Oscillations}\label{sec:dataBAO}

Acoustic sound waves propagated in the tighly coupled photo-baryon fluid before the decoupling of CMB photons at $z_*\simeq 1100$. They left an imprint in the distribution of galaxies that manifests itself as a peak in the two-point galaxy correlation function, which is located at the maximum distance traveled by the sound wave, i.e. the sound horizon at the baryon drag epoch, $r_d$. This peak translates into wiggles in the matter power spectrum, its Fourier transform. Several galaxy surveys have measured these features in the last twenty years with increasing degree of precision and spanning different redshift ranges \citep{2dFGRS:2005yhx,SDSS:2005xqv}. They use $r_d$ as a standard ruler with respect to which they measure cosmological distances at various redshifts. This can be employed to constrain cosmological models in a quite robust way \citep{Sherwin:2018wbu,Carter:2019ulk,Bernal:2020vbb,Brieden:2021edu,Brieden:2021cfg}. Their constraints are given either in terms of the dilation scale $D_V$,
\begin{equation}
\frac{D_V(z)}{r_d}=\frac{1}{r_d}\left[D_M^2(z)\frac{cz}{H(z)}\right]^{1/3}\,,
\end{equation}
or by splitting (when possible) the angular and radial BAO information, providing data on $D_{A}(z)/r_d$ and $H(z)r_d$ separately, with some degree of correlation. 

In any Riemannian metric theory of gravity with photons traveling on null geodesics and conservation of the photon number, the Etherington relation \citep{Etherington:1933} holds, 

\begin{equation}\label{eq:AngDista}
D_A(z)=\frac{D_L(z)}{(1+z)^2}\,.
\end{equation}
It is very useful, since it can be employed to convert angular diameter distances into luminosity distances, and vice versa. Current low-redshift data does not point to any deviation from this relation \citep{Renzi:2021xii}.

We show the list of BAO data points employed in this work and their corresponding references in Table \ref{tab:BAO_data}.


\section{Gaussian Processes}\label{sec:GP}

\subsection{The basics}\label{sec:basicsGP}
Data-driven reconstructions of cosmological functions subject to minimal model assumptions can be obtained with Gaussian Processes. Based on Bayesian statistics, this machine learning algorithm has become in recent years one of the most widely used model-independent regression techniques in cosmology. It requires the data to be Gaussianly distributed. 

A Gaussian Process 
$f(x) \sim {\rm GP} (\mu(x), D(x, \tilde x))$ is a generalization of a multivariate Gaussian, and is defined by the mean function $\mu(x)$ and the covariance matrix $D(x, \tilde x)$, see, e.g., \citep{2006gpml.book.....R}. If we denote the collection of the $n$ data points that will be employed to train the GP as $Y$, being the latter located at points $X$, the covariance matrix $D$ takes the following form
\begin{equation}\label{eq:covGP}
D(x,\tilde x)\equiv\Biggl \{
    \begin{array}{lcl}
        K(x,\tilde x) + C(x,\tilde x) & {\rm if}\,x\,{\rm and}\,\tilde{x} \in X\\
        K(x, \tilde x) & {\rm otherwise} \\
    \end{array}\,,
\end{equation}
where $C$ is the covariance matrix of the data and $K(x,\tilde x)$ the so-called kernel function. Imagine that we want to reconstruct our function at the locations $X^\star$ ($\ne X$). By computing the probability of finding a given realization
of the GP under the condition $f(X)=Y$ we find that the resulting GP is characterized by the mean function

\begin{equation}
    \bar f^{\star}=\mu^{\star}+ K(X^{\star},X)[K(X,X)+C(X,X)]^{-1}(Y-\mu)\,,
\end{equation}
and the covariance 


 \begin{equation}\label{eq:GPcov}
   {\rm cov}(f^{\star})= K(X^{\star},X^{\star})-K(X^{\star},X)[K(X,X)+C]^{-1}K(X,X^{\star})\,.
 \end{equation}
$\mu^{\star}\equiv\mu(X^\star)$ is the a priori assumed mean of the reconstructed function at $X^\star$. The kernel, which encodes the assumptions on the covariance between points at which we do not have data, plays a central role. There are many possible kernel functions to be employed in a GP, but the simplest choice falls into the category of stationary kernels, which depend only on the distance between the input data points, that is on $|x-\tilde{x}|$, and not on their individual values $x$ and $\tilde{x}$, being thus invariant to translations in the input space. Although the GP is regarded as a non-parametric method, the kernels introduce some hyperparameters that are typically in charge of controlling the strength of the fluctuations and the correlation length between two separate points. Before the reconstruction, these hyperparameters have to be determined by a proper optimization or marginalization of the GP. 
These two processes require the maximization or the sampling, respectively, of the likelihood

\begin{align}\label{eq:GPlike}
\ln \mathcal{L}=-\frac{1}{2}(Y-\mu)^{T}&[K(X,X)+C]^{-1}(Y-\mu)\nonumber\\
&-\frac{1}{2}\ln|K(X,X)+C|-\frac{n}{2}\ln(2\pi)\,,
\end{align}
which is obtained by marginalizing the GP over the points that are not contained in the data set. In many cases, this likelihood is sharply peaked and the optimized result becomes a good approximation \citep{Seikel:2012uu}. This is usually the case when a constant prior mean is employed in the analysis \citep{Hwang:2022hla}. However, strictly speaking, from a Bayesian perspective, getting the full distribution of the hyperparameters is the correct way to proceed. Indeed, if we want to take into account the correlations between the kernel hyperparameters and their uncertainties, we need to abandon the assumption that their distribution is a Dirac delta, see e.g. \citep{Gomez-Valent:2018hwc,Hwang:2022hla}. By doing so, the non-zero uncertainties of the hyperparameters can then be propagated to the reconstructed function under study. It is important not to neglect them or, at least, to duly assess their impact on the results. We will do so in Sec. \ref{sec:recHm}, together with a study of the impact of the prior mean $\mu$.

In this work we make use of the public package {\it Gaussian Processes in Python} (\texttt{GaPP}) \footnote{\url{https://github.com/carlosandrepaes/GaPP}}, first developed by \cite{Seikel:2012uu}. One of its modules is prepared to perform the Monte Carlo Markov Chain (MCMC) sampling of the kernel hyperparameters. It relies on the public package \texttt{emcee}\footnote{\url{https://emcee.readthedocs.io/en/stable/}} \citep{Foreman_Mackey_2013}, which is a \texttt{Python} implementation of the affine invariant MCMC ensemble sampler by \cite{Goodman2010}.

\begin{figure}    \centering
\includegraphics[width=\columnwidth]{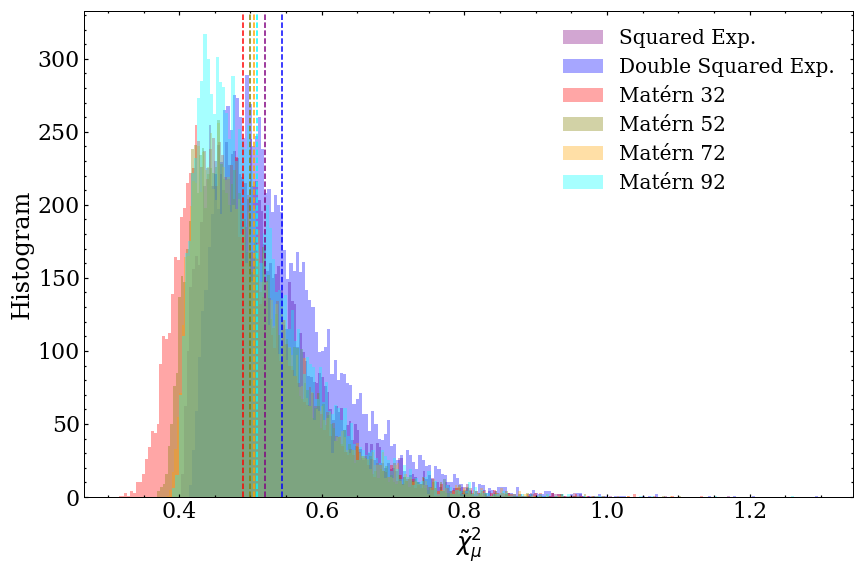}
    \caption{Histograms of the $\tilde \chi^{2}_\mu$ obtained for the reconstruction of $H(z)$ for the six kernels employed in the GP training (see Sec. \ref{sec:selectGP} for more details). The vertical dotted lines are located at the corresponding mean values. Notice that all of them are clearly below and far away from 1.}\label{fig:histcov}
\end{figure}


\subsection{A method to select the kernel}\label{sec:selectGP}

We now present a mathematical criterion to select the most suitable kernel among the available ones. We then apply it to the reconstruction of the Hubble function, $H(z)$.

There are six available kernels in the \texttt{GaPP} package. The simplest one is the Squared Exponential, defined as
\begin{equation}\label{eq:SE}
    K(x,\tilde x)=\sigma^{2}_{f}\exp\Bigl(-\frac{(x-\tilde x)^{2}}{2l^{2}}\Bigr)\,,
\end{equation}
where $l$ and $\sigma_f$ are two hyperparameters, in charge of controlling the correlation length between points and the amplitude of the uncertainties, respectively. The sum of two Squared Exponentials defines the so-called Double Squared Exponential kernel. In \texttt{GaPP} there are also four types of kernels contained in the Matérn family. If we define $\Gamma$ as the gamma function and $Y_{\nu }$ as the modified Bessel function of the second kind, the Matérn covariance between two points separated by the distance $d=|x-\tilde x|$ is given by

\begin{equation}\label{eq:matern}
    {\displaystyle K_{\nu }(d)=\sigma_{f} ^{2}{\frac {2^{1-\nu }}{\Gamma (\nu )}}{\Bigg (}{\sqrt {2\nu }}{\frac {d}{l }}{\Bigg )}^{\nu }Y_{\nu }{\Bigg (}{\sqrt {2\nu }}{\frac {d}{l }}{\Bigg )}}\,,
\end{equation}
where $\nu =p+1/2,\ p\in \mathbb {N}^{+}$. In the limit $\nu\to\infty$ we recover the Squared Exponential kernel. The Matérn covariance family is $m$ times differentiable in the mean-square sense, i.e. the derivative $\partial^{2m}K(x,\tilde x)/\partial z^{m}\partial\tilde z^{m}$ exists and is finite if $\nu>m$. Higher values of $\nu$ translate into wider peaks and smoother reconstructed functions due to the stronger correlation between points. \texttt{GaPP} contains the Matérn kernels with $\nu=3/2,5/2,7/2,9/2$, called Matérn 32, 52, 72 and 92, respectively. 
 
We perform the reconstruction of $H(z)$ employing the 32 CCH data points listed in Table \ref{tab:CCH} with the GP trained with the six aforementioned kernels in the redshift range $0\leq z\leq 1.965$. We show the results obtained from each kernel in Appendix \ref{sec:AppendixA}, see Fig. \ref{fig:kernelsperf}. Not very significant differences can be appreciated between them with naked eye. To assess the performance of the kernels in the reconstruction of $H(z)$, we proceed as follows. We draw with each kernel $N_{real}=10^{4}$ GP random realizations, $H_{rec_\mu}(z)$ with $\mu=1,...,N_{real}$, accounting for both the covariance of the data points and of the reconstruction. For each realization we compute the $\chi^{2}$ statistics, using the following expression,

\begin{table}
    \centering
    \caption{Results of the test based on the $\tilde \chi^{2}_{\mu}$ analysis to determine the kernel that performs the best for the reconstruction of $H(z)$. In the first column we indicate the pairs of kernels under comparison, taking in all cases the Squared Exponential (SE) as reference. We use the following shorthand notation: Double Squared Exponential (DSE), Matérn 32 (M32), Matérn 52 (M52), Matérn 72 (M72), and Matérn 92 (M92). The second column shows the relative weight of the kernels. The best-performing kernel is Matérn 32, cf. the line in bold and Sec. \ref{sec:selectGP} for more details.}
    \label{tab:kernel_perform}
    \begin{tabular}{c|c}
        \hline
        Kernels & $P_{\rm SE}/P_j$   \\
        \hline
        SE vs DSE  & 1.42  \\ 
        \bf SE vs M32 &   \bf 0.62  \\
        SE vs M52 & 0.72   \\
        SE vs M72  & 0.82  \\
        SE vs M92   &0.83   \\
        \hline
    \end{tabular}
\end{table}

\begin{equation}\label{eq:kernel_chi2}
\chi_{\mu}^{2}=\sum_{i,j=1}^{32} [H(z_i)-H_{rec,\mu}(z_i)]\tilde{C}_{ij}^{-1}[H(z_j)-H_{rec,\mu}(z_j)]\,,
\end{equation}
where $\tilde{C}$ is the covariance matrix of the CCH data and the Latin indices label the $n_p=32$ redshifts at which we have data. Thus, the $N_{real}$ realizations of the Hubble function lead to $N_{real}$ values of $\chi^{2}_\mu$. More concretely, in order to penalize the use of additional hyperparameters, we compute the reduced $\chi^2$, $\tilde \chi^{2}_\mu= \chi^{2}_\mu/{\rm dof}$, with dof being the number of degrees of freedom, i.e. the number of data points minus the number of hyperparameters. We then build a histogram of $\tilde \chi^{2}_\mu$ for each kernel, cf. Fig. \ref{fig:histcov}. Several comments are in order. First, the figure shows that the mean values of $\tilde{\chi}^{2}$ lie below and quite far from 1, regardless of the kernel. This might be due to an overestimation of the CCH uncertainties. In Sec. \ref{sec:spec} we will speculate about this possibility and see how our results change when we allow the CCH data to take smaller errors. Secondly, the kernel Matérn 32 is the one with the lowest mean $\tilde{\chi}^{2}$. However, we need to estimate more quantitatively the relative ability of the kernels to describe the data. Let us consider two kernels $K_i$ and $K_j$. The probability that the reduced $\tilde \chi^{2}_\mu$ associated to $K_i$ is lower than the one associated to $K_{j}$ reads,

\begin{equation}\label{eq:P_ratio}
    P_{\tilde \chi^{2}_{K_{i}}<\tilde \chi^{2}_{K_{j}}}=\frac{1}{1+P_{j}/P_{i}}\,,
\end{equation}
where $P_{j}/P_{i}$ in the right-hand side is the ratio of their statistical weights. In practice, if we use a sufficiently large number of realizations, $N_{real}$, we can estimate $P_{j}/P_{i}\sim N_j/N_i$, with $N_i$ being the number of realizations in which $\tilde \chi^{2}_{K_{i}}<\tilde \chi^{2}_{K_{j}}$ and $N_j=N_{real}-N_i$. As we are computing relative weights, we can set e.g. $j=1$, and compute its relative performance with respect to the Kernels $K_2,..,K_6$. In the analysis presented in Table \ref{tab:kernel_perform} $K_1$ stands for the Squared Exponential kernel. It is clear from that table that Matérn 32 is the best-performing kernel regarding the reconstruction of $H(z)$. For completeness, we also check whether this result is sensitive to the ordering of the vectors containing the values of $\tilde{\chi}^2_\mu$. The results are very stable. Indeed, the ratios $P_{\rm SE}/P_i$ differ only by a tiny percentage, which is only due to numerical noise, i.e. it becomes smaller and smaller for increasing values of $N_{real}$.


\begin{figure}
    \centering
    \includegraphics[width=\columnwidth]{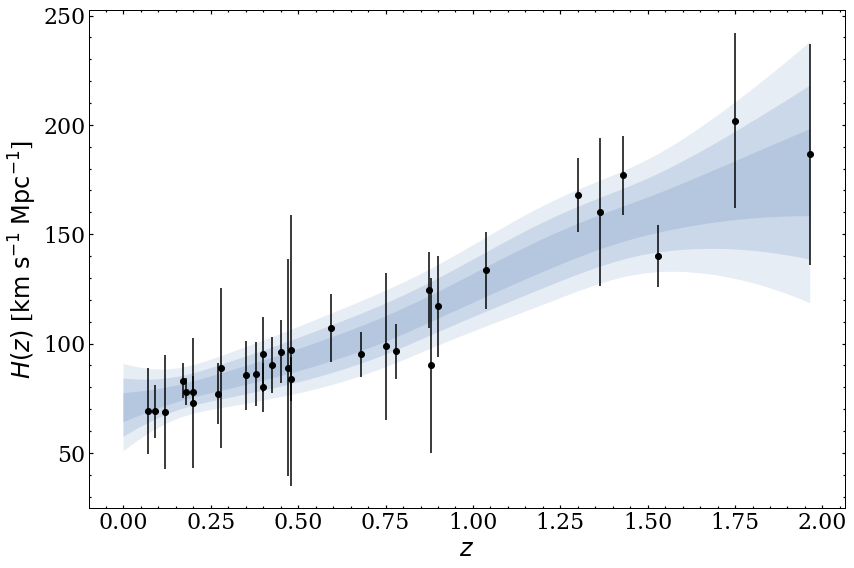}
    \includegraphics[width=\columnwidth]{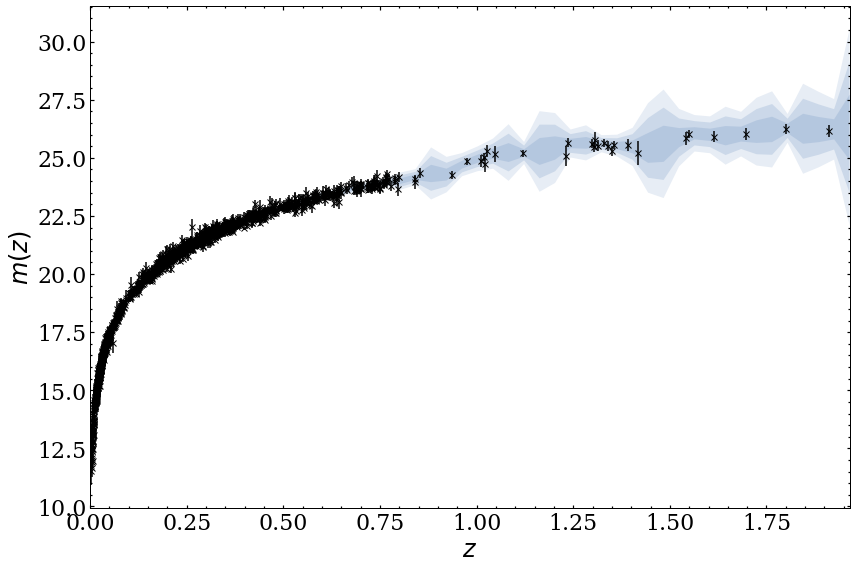}
    \caption{\textit{Upper plot:} Reconstructed shape of the Hubble function $H(z)$ at 1$\sigma$, 2$\sigma$ and 3$\sigma$ obtained from Gaussian Processes and the CCH data of Table \ref{tab:CCH} (in black). \textit{Lower plot:} The same, but for the apparent magnitude of SNIa, $m(z)$ Eq. \eqref{eq:mfinal}. In this case we use the observational data from the Pantheon+ compilation, as explained in Sec. \ref{sec:dataSNIa}. See Sec. \ref{sec:GP} for details about the GP method.}\label{fig:GP_rec}
\end{figure}

\subsection{Reconstruction of \texorpdfstring{$H(z)$}{} and \texorpdfstring{$m(z)$}{}}\label{sec:recHm}

We reconstruct now the shape of the Hubble function $H(z)$ and the apparent magnitude of SNIa $m(z)$ using Gaussian Processes and the data described in Secs. \ref{sec:dataCCH} and \ref{sec:dataSNIa}, respectively. As anticipated in the Introduction, the aim of obtaining these model-independent reconstructions is to use them (among other things) to reconstruct first the absolute magnitude of SNIa and the curvature parameter as a function of the redshift, see Sec. \ref{sec:tests}.

We obtain $H(z)$ from CCH following the method and the prescriptions described in Secs. \ref{sec:basicsGP} and \ref{sec:selectGP}, i.e. using the  Matérn 32 kernel, a zero mean function $\mu$, and taking into account the full distribution of the hyperparameters $\sigma_f$ and $l$. This, in particular, only has a modest impact on the final reconstruction. The mean in the marginalization procedure differs by 5\% at most from the optimized result and the errors are 8\% larger. Moreover, we have explicitly checked that we obtain very similar results using $\mu=0,10,100$. They differ only by $\lesssim 0.1\sigma$. Hence, a full marginalization process that includes also the marginalization over a constant $\mu$ (together with the hyperparameters) leads essentially to the same final reconstructed shape of $H(z)$. In addition, we have also studied what happens if we assume a prior mean based on the $\Lambda$CDM prediction, marginalizing also over the parameters $H_0$ and $\Omega_m$. We find that this introduces very strong model dependencies, basically yielding the same output as in a pure $\Lambda$CDM fit. This goes against the philosophy of our work, so we prefer to use a constant mean in our main analyses.

Due to the large covariance matrix of the Pantheon+ compilation, it is very expensive from the computational point of view to perform the marginalization over the hyperparameters and repeat the analysis of Sec. \ref{sec:selectGP} for $m(z)$, so we opt to use also in this case the Matérn 32 kernel and the best-fit values obtained from the maximization of the marginalized likelihood Eq. \eqref{eq:GPlike}. Using the binned Pantheon data from \citep{Pan-STARRS1:2017jku}, we have checked that the results are not very sensitive to these choices. In addition, we employ the reconstruction of $m(z)$ only in some of the tests of Sec. \ref{sec:tests}. The conclusions of these tests do not depend on these subtleties. To obtain the final constraints on the triad of parameters $(M,\Omega_k,r_d)$ in Sec. \ref{sec:CalibSec} we only make use of the reconstruction of the Hubble rate, which duly incorporates the uncertainties of the hyperparameters.  

We show the reconstructed shapes of $H(z)$ and $m(z)$ in Fig. \ref{fig:GP_rec}. The extrapolated value of the Hubble parameter reads, $H_0=(70.7\pm 6.7)$ km/s/Mpc. For previous reconstructions of the Hubble rate with GPs and CCH see e.g. \citep{Busti:2014dua,Yu:2017iju,Gomez-Valent:2018hwc,Haridasu:2018gqm,Yang:2022jkf,Renzi:2020fnx}, and for previous reconstructions of $m(z)$ or the distance modulus from SNIa data see e.g. \citep{Seikel:2012uu,Cai:2015pia,Yu:2016gmd,Yang:2020bpv,Liang:2022smf,Renzi:2020fnx}.

\section{Some tests of the consistency of low-\texorpdfstring{$z$}{} data and the theoretical assumptions behind the standard cosmological model} \label{sec:tests}

\subsection{Testing the constancy of \texorpdfstring{$M$}{} }\label{sec:testM}

\begin{figure}
    \centering
    \includegraphics[width=\columnwidth]{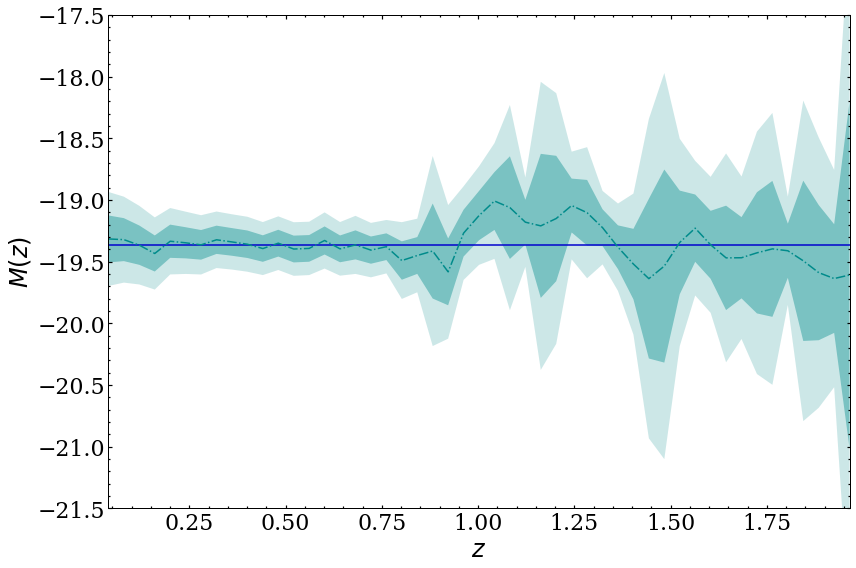}
    \includegraphics[width=\columnwidth]{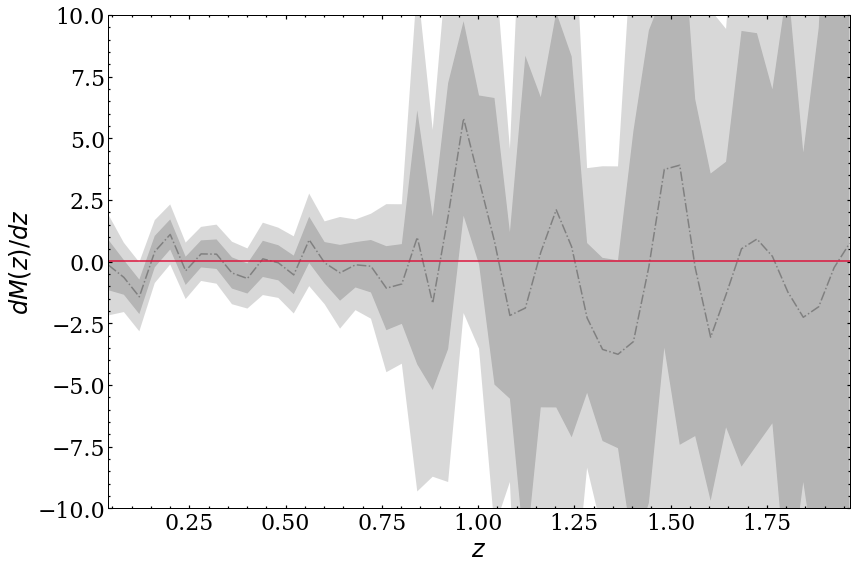}
    \caption{Reconstruction of $M(z)$ and its derivative $dM/dz$ at 68\% and 95\% C.L., obtained by fixing $\Omega_k=0$ and using the method described in Sec. \ref{sec:testM}. The constant lines appearing in the plots (in blue and red, respectively) are the corresponding  weighted means, computed with formula \eqref{eq:meanM}.}
    \label{fig:M_reco}
\end{figure}

In this section we reconstruct the shape of the absolute magnitude of SNIa, $M(z)$, in order to test its constancy throughout the cosmic expansion, in the redshift range $z\lesssim 2$. In Sec. \ref{sec:recHm} we have obtained the GPs associated to $H(z)$ and $m(z)$. We can reconstruct $M(z)$ using formula \eqref{eq:mfinal}. First, we draw $N_{real}=5\cdot10^{5}$ realizations (curves) of the Hubble function and the apparent magnitude from their corresponding Gaussian Process. In order to compute the reconstructed shape of the luminosity distance using Eq. \eqref{eq:DL} we have to employ a prior for the curvature parameter. For this first analysis we fix $\Omega_k=0$, while then we will study also other values to assess the impact of this prior on our result. For each realization in the sample, we compute $M(z_i)$ at $n_p=50$ equispaced knots. We present the reconstructed shape of $M(z)$ and its first derivative, $dM/dz$, in Fig. \ref{fig:M_reco}. From these plots it is evident that the resulting function is fully compatible with a constant at $\lesssim 1\sigma$. We have checked that this statement actually holds for a wide range of values of the curvature parameter $\Omega_k\in[-1,1]$.

In view of these results, it is natural to estimate the value of the constant $M$ from the reconstructed shape of $M(z)$. As we have $n_p$ knots we have $n_p$ distributions of $M$, i.e. one for each knot. These samples are correlated, of course. Assuming that they are Gaussianly distributed, we can construct a probability distribution that takes the following form,

\begin{equation}\label{eq:likeM1}
\mathcal{L}(M)=\mathcal{N}\exp\left[-\frac{1}{2}\sum_{i,j=1}^{n_p}(M-\bar{M}_i)(M-\bar{M}_j)(C^{-1})_{ij}\right]\,,
\end{equation}
with $\mathcal{N}$ the normalization constant and $\bar{M}_{i}$ the mean value in the $i$-th knot. Let us define now $A_{ij}\equiv (C^{-1})_{ij}$ to simplify the notation\footnote{Notice that this covariance matrix $C$ is different to the one defined in the preceding formula \eqref{eq:covM}.}. It is easy to show that Eq. \eqref{eq:likeM1} can be rewritten as follows

\begin{equation}\label{eq:likeM2}
\mathcal{L}(M)=\tilde{\mathcal{N}}\exp\left[-\frac{1}{2}\left(\sum_{i,j=1}^{n_p} A_{ij}\right)\left(M-\frac{\sum\limits_{i,j=1}^{n_p} \bar{M}_iA_{ij}}{\sum\limits_{i,j=1}^{n_p} A_{ij}}\right)^2\right]\,.
\end{equation}
This means that the distribution of $M$ is a Gaussian with mean and deviance

\begin{equation}\label{eq:meanM}
\bar{M} = \frac{\sum\limits_{i,j=1}^{n_p} \bar{M}_iA_{ij}}{\sum\limits_{i,j=1}^{n_p} A_{ij}}\qquad ;\qquad \sigma^2=\frac{1}{\sum\limits_{i,j=1}^{n_p} A_{ij}}\,,
\end{equation}
respectively. We estimate the covariance matrix from our sample as follows,

\begin{equation}
C_{ij}=\frac{1}{N_{real}}\sum_{\mu=1}^{N_{real}}(M_{\mu,i}-\bar{M}_i)(M_{\mu,j}-\bar{M}_j)\,,   
\end{equation}
where $M_{\mu,i}$ is the value of the absolute magnitude at the $i$-th knot for each realization $\mu=1,...,N_{real}$.

\begin{figure}
    \centering    \includegraphics[width=\columnwidth]{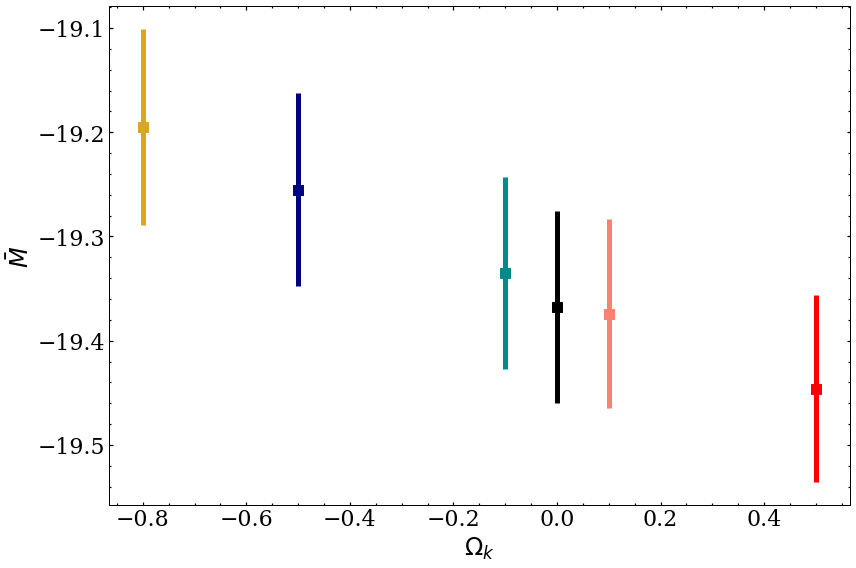}
    \caption{Constraints at $1\sigma$ C.L. on the constant values of $M$ obtained from the reconstructed shapes of $M(z)$ with different priors on $\Omega_k$. They are taken to be Dirac deltas located at the points of the $x$-axis. The shapes of $M(z)$ are in all cases consistent with constant values (see Fig. \ref{fig:M_reco} and Sec. \ref{sec:testM}), but it is clear from this plot that these constants are strongly dependent on the prior we use for the curvature. In order to get consistent constraints for both, $M$ and $\Omega_k$, we need to perform a joint analysis. See Sec. \ref{sec:CalibSec}.}
    \label{fig:prior_impact}
\end{figure}

Applying these formulas, we find $\bar{M}=(-19.368\pm0.092)$ mag in the case in which we set $\Omega_k=0$. However, this result can only be considered as a first approximation for two reasons: (i) non-Gaussian features, despite being small, can introduce some mild changes, which are not captured by the distribution Eq. \eqref{eq:likeM1}. However, we have performed a sanity check to verify that the $n_{p}$ distributions of $M$ are Gaussian in very good approximation. At each redshift point where we reconstruct $M(z)$, we build a histogram from its $N_{real}$ realizations and check that the skewness of each of them is compatible with zero. Hence, the bias introduced by this fact is certainly very small; and (ii) in this calculation all the redshift range is equally weighted, but in reality the data points are not uniformly distributed and this might also have an impact on the estimation of the weighted mean and its uncertainty. 

\begin{figure*}
    \centering
    \includegraphics[width=\columnwidth]{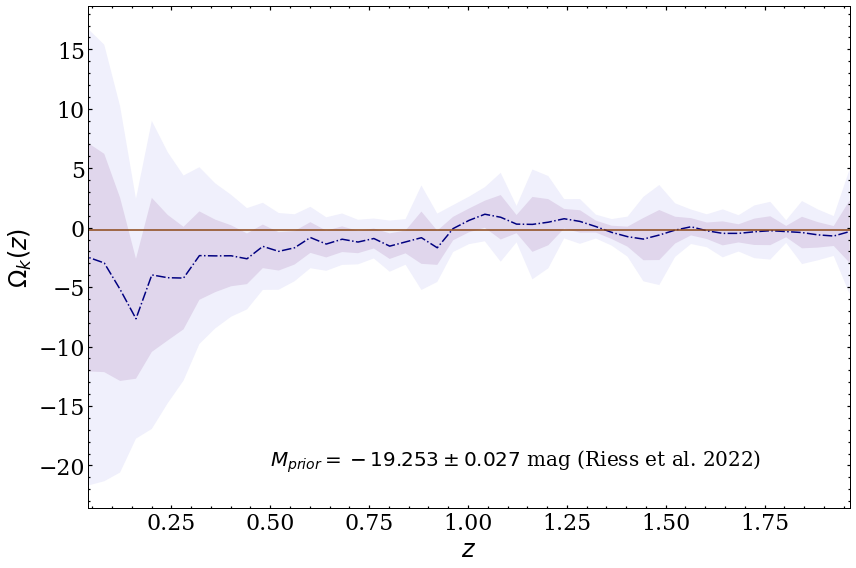}
    \includegraphics[width=\columnwidth]{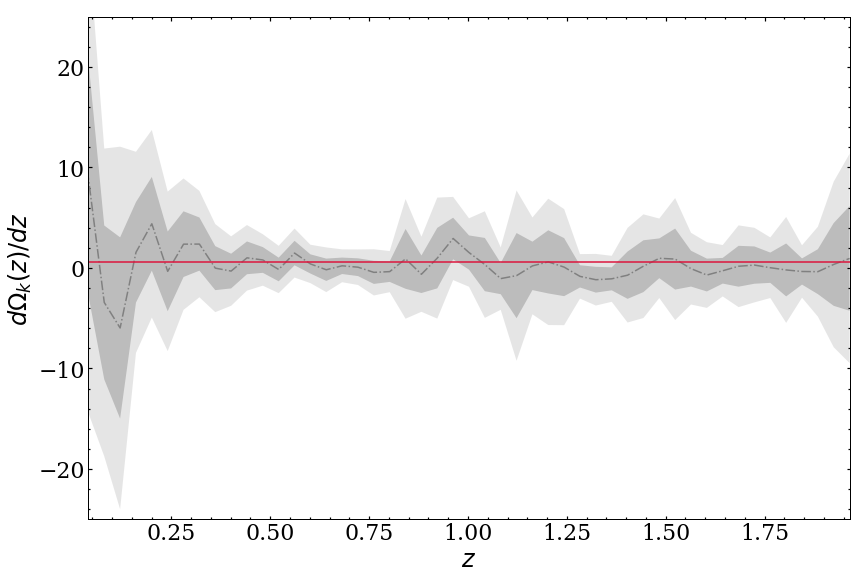}
    \includegraphics[width=\columnwidth]{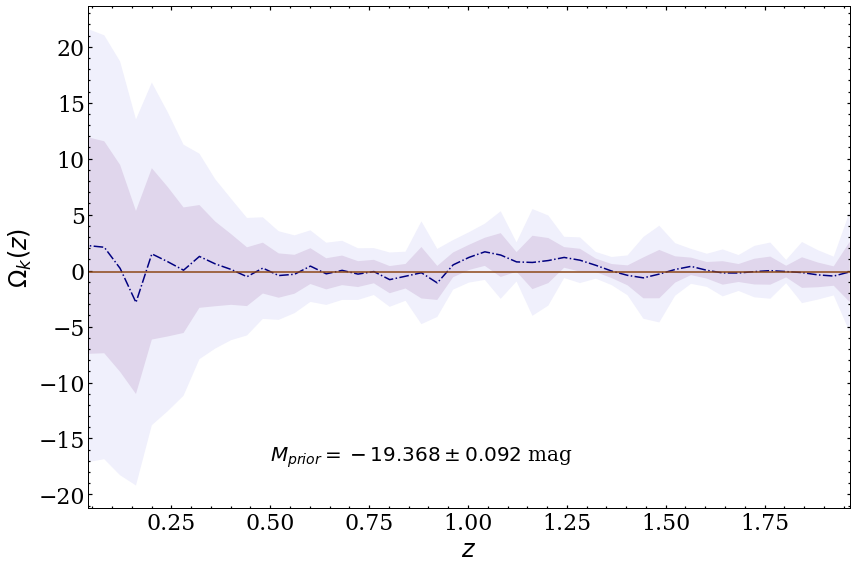}
    \includegraphics[width=\columnwidth]{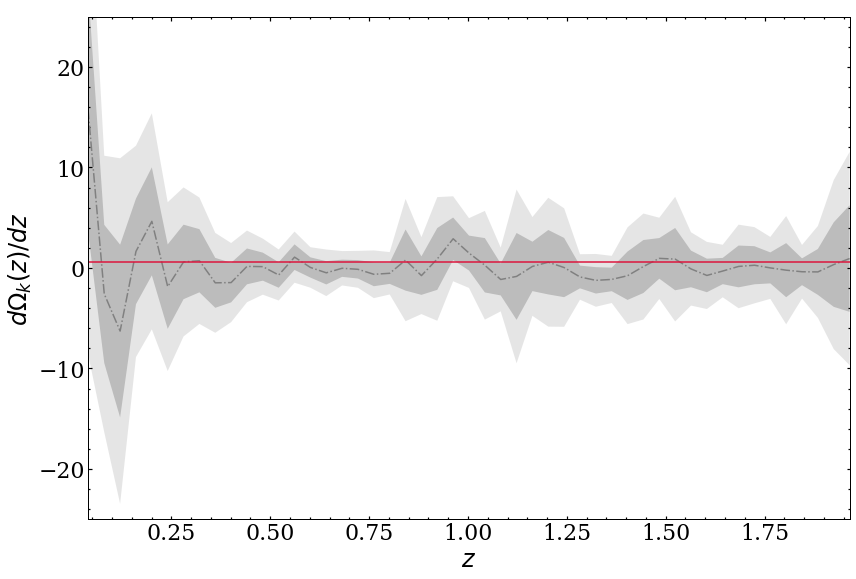}
    \caption{\textit{Upper plots:} Reconstructed shape of $\Omega_k(z)$ and $d\Omega_k(z)/dz$ at 1$\sigma$ and 2$\sigma$ C.L. obtained as explained in Sec. \ref{sec:testCP}, assuming the SH0ES prior $M^{R22}$ \citep{Riess:2021jrx}.  \textit{Lower plots:} The same, but using the Gaussian prior for $M$ obtained from the analysis presented in Fig. \ref{fig:M_reco}, $M=(-19.368\pm 0.092)$ mag.  
    In all the plots the dot-dashed lines correspond to the mean curves of $\Omega_k(z)$ and $d\Omega_k(z)/dz$, while the constant solid lines are their overall weighted mean, computed again with formula \eqref{eq:meanM}. We get the mean values $\bar{\Omega}_k=-0.21$ and $\bar{\Omega}_k=-0.17$ in the upper and lower plots, respectively.}
    \label{fig:omegak_Sprior}
\end{figure*}

As mentioned before, the shape of $M(z)$ is compatible with a constant regardless of the value of $\Omega_k$ chosen to carry out the analysis. Nevertheless, it is important to notice that the value of that constant depends a lot on the prior. In Fig. \ref{fig:prior_impact} we show how the constraint on $M$ changes with $\Omega_k$, from values of $M\sim -19.2$ mag to $M\sim-19.45$ mag when $\Omega_k$ varies from $-0.8$ to $+0.5$. The range of values of $\Omega_k$ explored here is much broader than what is allowed by the $\Lambda$CDM constraints from {\it Planck} \citep{Aghanim:2018eyx}. This has to be consistent with our model-independent approach. As we will see in Sec. \ref{sec:CalibSec}, large absolute values of the curvature are not excluded by the low-redshift data sets employed in this paper.

The test done in this section demonstrates that with CCH and Pantheon+ data sets, there is no significant statistical preference for the evolution of $M(z)$. However, if we want a robust estimate on the constant value of $M$, we need to constrain simultaneously both $M$ and $\Omega_k$ in a joint analysis. This will become even more evident in Sec. \ref{sec:testCP}, where we reconstruct $\Omega_k(z)$.


\subsection{Testing the Cosmological Principle} \label{sec:testCP}

Now, we reconstruct $\Omega_k(z)$. The result can be employed to test the Cosmological Principle without specifying the energy content of the universe nor the gravity action. \cite{Clarkson:2007pz} proposed to use 

\begin{equation}\label{eq:ClarksonExp}
\Omega_k(z) = \frac{[H(z)D^\prime_M(z)/c]^2-1}{[H_0D_M(z)/c]^2}\,,
\end{equation}
with the prime denoting a derivative with respect to the redshift, as a diagnostic of the homogeneity of the universe. This expression is obtained straightforwardly from Eq. \eqref{eq:DL}. Deviations of it from a constant value at any redshift can be considered to be a hint of the breaking of the CP. The function \eqref{eq:ClarksonExp} can be reconstructed from measurements of $H(z)$ and the luminosity distance. Hence, we can build it from CCH and calibrated SNIa data.

Here, we reconstruct the curvature parameter as a function of $z$, but in an alternative way, which allows us to skip the numerical computation of the derivatives $D_M^\prime(z)$ appearing in Eq. \eqref{eq:ClarksonExp}. It works as follows. We use the GP of $H(z)$ to generate $N$ samples of the Hubble function. On the other hand, we draw $N$ values of $M$ from the SH0ES Gaussian prior on the absolute magnitude of SNIa, $M^{R22}$. With the latter and $N$ GP-realizations of $m(z)$ we can reconstruct $D_L(z)$ using formula \eqref{eq:mfinal}, and also the angular diameter distance through the Etherington relation \eqref{eq:AngDista}. We employ all these ingredients to solve Eq.  \eqref{eq:DL} numerically for every redshift and find $N$ realizations of $\Omega_k(z)$. Our results are presented in Fig. \ref{fig:omegak_Sprior}. The reconstructed function is compatible with a constant, so there is no hint of a violation of the CP. This resonates well with previous results in the literature obtained with older data sets and applying a different methodology, see e.g. \citep{Cai:2015pia,Yu:2016gmd,Yang:2020bpv}. We have verified that this finding is again independent of the prior on $M$ employed in the analysis, although the constraint we get on the constant value of $\Omega_k$ does depend on it. This is evident from Fig. \ref{fig:omegak_Sprior}, see also the caption. In Sec. \ref{sec:CalibSec} we will provide joint constraints on $M$ and $\Omega_k$ in order to get rid of the ambiguity introduced by the subjective choice of the priors.


\begin{figure}
    \centering
    \includegraphics[width=\columnwidth]{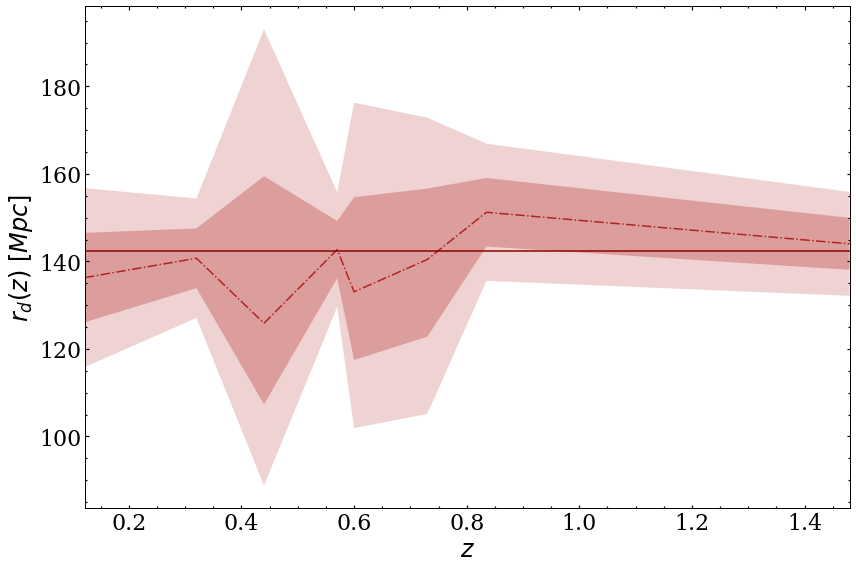}
    \includegraphics[width=\columnwidth]{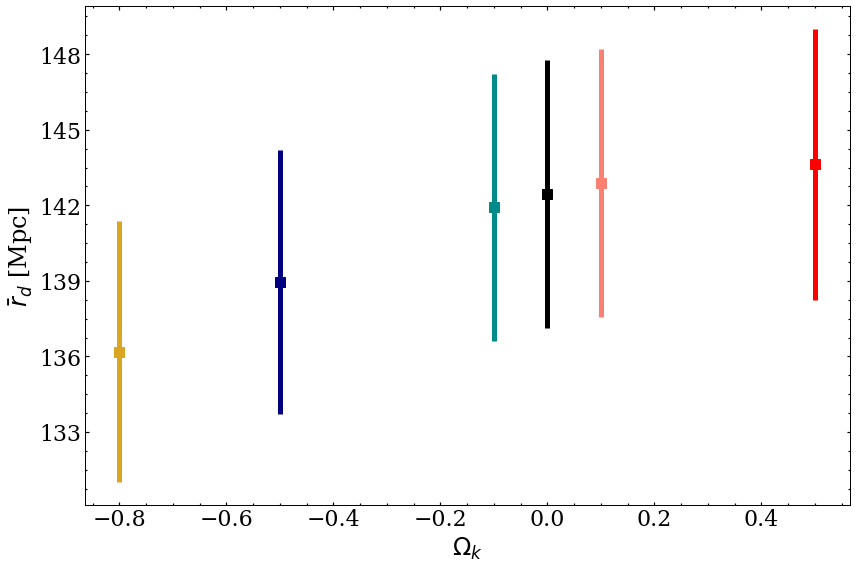}
    \caption{{\it Upper plot:} Result of the consistency test of the BAO data listed in Table \ref{tab:BAO_data}, as described in Sec. \ref{sec:testBAO}. The plot shows measurements of $r_d$ (at $68\%$ and $95\%$ C.L.) as a function of the redshift, fixing $\Omega_k=0$. The dot-dashed line passes through the peak values of the individual histograms of $r_d(z)$ at each redshift, while the solid line represents the weighted mean obtained from formula \eqref{eq:meanM}, which reads: $\bar{r}_d= (142.5\pm5.3)$ Mpc. {\it Lower plot:} Dependence of the weighted mean $\bar{r}_d$ (at 
$1\sigma$ C.L.) on the prior value of $\Omega_k$ employed in the analysis.}
    \label{fig:BAO_consistency}
\end{figure}

\subsection{Testing the consistency of the BAO data} \label{sec:testBAO}

In this section we test the internal robustness of the BAO data listed in Table \ref{tab:BAO_data} in the light of the CCH data. Given the reconstructed expansion rate derived from CCH, we would expect the values of $r_{d}$ obtained from the various BAO data points to be statistically consistent with each other. Otherwise, this could signal the presence of uncorrected systematic effects in the data.

We apply a method that is completely analogous to the one performed to obtain $M(z)$ and $\Omega_k(z)$ in Secs. \ref{sec:testM} and \ref{sec:testCP}, respectively. We use the GP for $H(z)$ to generate curves of the Hubble function. From them we can also reconstruct $D_{A}(z)$ for a fixed curvature parameter (we first consider the case of a flat universe, i.e. $\Omega_k=0$). We compute the angular diameter distances and $H(z)$ at the redshifts of the BAO data points. We then draw Gaussian-distributed vectors of BAO data and combine this information to obtain 11 distributions of $r_{d}$, i.e. one for each BAO data point. 

We present our results in the upper plot of Fig. \ref{fig:BAO_consistency}. For those redshifts with two BAO data points (at $z=0.32, 0.57, 1.48$, cf. Table \ref{tab:BAO_data}) we use the weighted mean and uncertainty as provided in formula \eqref{eq:meanM} to obtain a single value of $r_d$. It is clear from that plot that the values of $r_d$ at the various redshift values are consistent with each other. This result still holds (at $68\%$ C.L.) if we allow the universe to take closed or open geometries, as we have explicitly checked by exploring values of $\Omega_k\in[-0.8,+0.5]$.


\section{Calibration of the cosmic ladders and measurement of \texorpdfstring{$\Omega_k$}{}}\label{sec:CalibSec}

The analyses carried out in Sec. \ref{sec:tests} show no evidence for an evolution of the absolute magnitude of SNIa with redshift nor a departure from homogeneity at large scales. Moreover, we have checked that the BAO data employed in this work are consistent and lead to values of $r_d$ that are fully compatible with each other. Hence,  we are legitimated to perform an analysis to jointly constrain the curvature parameter and the calibrators of the distance ladders by treating them simply as constants\footnote{We still assume cosmological isotropy, even though this symmetry of the CP has not been tested by us. See \citep{Aluri:2022hzs} for a review of the CP and hints for deviations from it.}.

\begin{figure*}
    \centering
    \includegraphics[width=5.5in, height=5.5in]{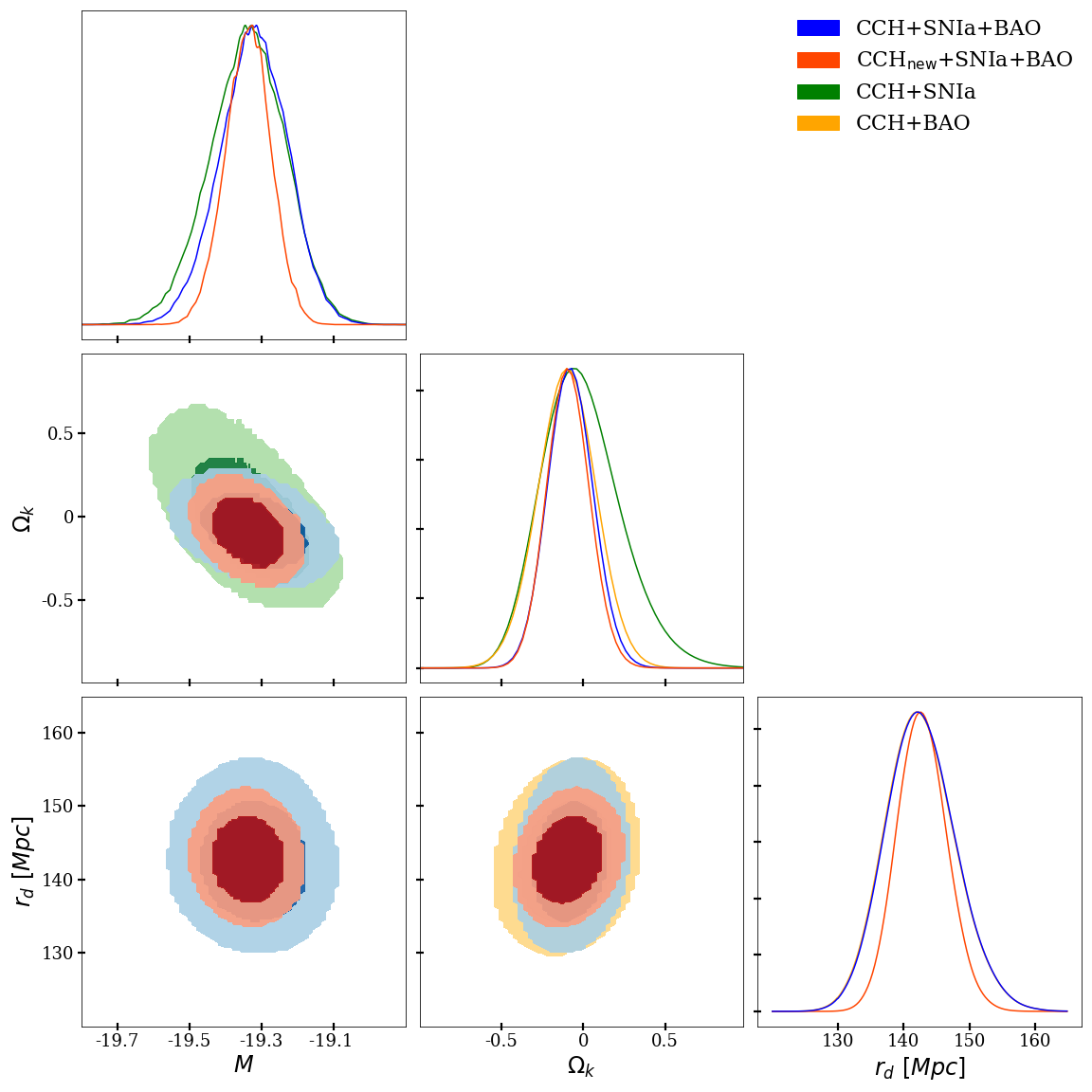}
    \caption{Two-dimensional contour plots in all the planes of the parameter space ($M, \Omega_k, r_d$) and the corresponding one-dimensional posterior probability distributions obtained from the joint analyses described in Secs. \ref{sec:CCH+SN}-\ref{sec:spec}. The 2D contours are evaluated at 68\% and 95\% C.L. As expected, the combination of the three data sets (CCH+SNIa+BAO) permits to obtain tighter constraints with respect to the CCH+SNIa and CCH+BAO analyses (see the estimated values in Table \ref{tab:joint}). This trend is even more remarkable if we allow uncertainties on the CCH data to decrease by a factor $\sim3/2$ (in red). See Sec. \ref{sec:spec} for more details.}
    \label{fig:joint3D}
\end{figure*}

We obtain constraints in the planes $(M,\Omega_k)$ and $(r_d,\Omega_k)$ using CCH+SNIa (in Sec. \ref{sec:CCH+SN}) and CCH+BAO (in Sec. \ref{sec:CCH+BAO}), respectively, making use of a quite model-independent approach, which is also independent of the data sets that drive the $H_0$ tension. Both, uncalibrated SNIa and BAO, are relative distance indicators. In practice, we use CCH to calibrate the standard candles and the standard rulers. Finally, in Sec. \ref{sec:CCH+SN+BAO} we combine the three data sets CCH+SNIa+BAO to constrain the full parameter space $(M,r_d,\Omega_k)$. The results of these analyses are shown in Fig. \ref{fig:joint3D} and the derived constraints on the various parameters are presented in Table \ref{tab:joint}.

In Sec. \ref{sec:SNIahost} we include the SNIa in the host galaxies and the information of their distances (inferred from calibrated Cepheid variable stars) to assess their impact. In Sec. \ref{sec:spec} we speculate about the possibility that uncertainties on the CCH have been overestimated. Specifically, we study a case in which CCH uncertainties have been lowered to get a distribution of $\tilde{\chi}_\mu^2$ with a mean equal to one (see Sec. \ref{sec:selectGP}). 

\begin{table}
    \centering
    \caption{Constraints on $M$, $\Omega_k$ and $r_d$ obtained from the joint analyses of Secs. \ref{sec:CCH+SN}-\ref{sec:CCH+SN+BAO}, see also Fig. \ref{fig:joint3D}. We remind the reader that these results are independent from the SNIa calibration with Cepheids in the first rungs of the cosmic distance ladder, since we exclude the SNIa contained in the host galaxies employed by SH0ES in their analysis \citep{Riess:2021jrx,Scolnic:2021amr}. Notice that the central values for $\Omega_k$ obtained with CCH+SNIa and CCH+SNIa+BAO are the same and a bit larger than the one obtained with CCH+BAO. In reality, the CCH+SNIa+BAO constraint for $\Omega_k$ falls somewhere in the middle between the CCH+SNIa and CCH+BAO results, as expected, but we are limited by the resolution of our grid, since we use a step $\Delta \Omega_k=0.03$. In any case, this step is much lower than the uncertainty of $\Omega_k$, so this fact has no impact on our conclusions.}
    \label{tab:joint}
    \begin{tabular}{cccc}        
         & CCH+SNIa & CCH+BAO & CCH+SNIa+BAO \\ 
        \hline
        \hline
        M [mag]& $-19.344^{+0.116}_{-0.090}$ &  & $-19.314^{+0.086}_{-0.108}$  \\       
         \hline
         $\Omega_k$ & $-0.07^{+0.27}_{-0.21}$ & $-0.10\pm0.18$ & $-0.07^{+0.12}_{-0.15}$ \\  
         \hline     
         $r_d$ [Mpc] & & $141.9^{+5.6}_{-4.9}$& $142.3\pm5.3$ \\        
        \hline
    \end{tabular}
\end{table}

\subsection{Analysis with CCH+SNIa}\label{sec:CCH+SN}
We employ the CCH and SNIa data sets to obtain joint constraints in the plane $(M,\Omega_k)$ making use of a grid-search method. First, we employ the GP trained with the CCH data to get $N=1.5\cdot10^{6}$ reconstructed curves of $H(z)$, from which we obtain $N$ reconstructions of $I(z)=\int^{z}_{0}dz^{\prime}/H(z^{\prime})$, i.e. the integral that enters the expression of the luminosity distance Eq. \eqref{eq:DL}. Actually, we only need to keep the values of this function at the redshifts at which we have the SNIa data, so we end up with $N$ vectors of values of $I(z)$.  Then, we build a rectangular grid in the plane ($M, \Omega_k$), with $M \in [-19.8, -18.9]$ mag and $\Omega_k \in [-1,1]$. The size of the steps is not uniform, we use smaller steps in those regions of the plane with a higher probability. This determines the total number of points that make up our grid. At each point of the grid, which is characterized by the values of $M$ and $\Omega_k$, we transform the $N$ vectors with $I(z)$ into $N$ vectors with $D_L(z)$ by virtue of Eq. \eqref{eq:DL} and, subsequently, in $N$ vectors with 
 the apparent magnitude $m_{rec}(z)$. This enables us to perform a $\chi^2$ analysis using the SNIa data. For each $i=1,..,N$ realizations in the $\mu$-th knot, we have
 
\begin{equation}\label{eq:chi2_mu}
    \chi_{\mu,i}^{2}=\sum_{k,l=1}^{1624}[{m}(z_k)-{m}_{rec,\mu,i}(z_k)]C_{kl}^{-1}[{m}(z_l)-{m}_{rec,\mu,i}(z_l)]\,,
\end{equation}
where $C$ is here the covariance matrix of the SNIa.

To evaluate the behaviour of $M$ and $\Omega_k$ and constrain these parameters, we can now use an estimator, $\chi^{2}_{eff}$, which associates
at each knot of the grid a weight $w_\mu$ proportional to
\begin{equation}
w_\mu\propto B_\mu \sum_{i=1}^{N} \exp(-\chi_{\mu,i}^2/2)\,.
\end{equation}
where the factor $B_{\mu}=\Delta M \cdot \Delta \Omega_k |_{\mu}$ accounts for the size of the bins at the $\mu$-th knot. We use flat priors for $M$ and $\Omega_k$. We can also rewrite the last expression in a slightly different way in order to ease its numerical computation,

\begin{align}\label{eq:w_mu}
w_\mu\propto & B_\mu\exp(-\bar{\chi}^2_{\mu}/2) \underbrace{\sum_{i=1}^{N} \exp(-[\chi_{\mu,i}^2-\bar{\chi}^2_{\mu}]/2)}_{\equiv f_\mu}\,,\nonumber\\
w_\mu\propto&B_\mu\exp(-\bar{\chi}^2_{\mu}/2) f_\mu\,,
\end{align}
with $\bar{\chi}^2_{\mu}$ the mean of the $\chi^2$ in that particular knot.
Our estimator reads, 

\begin{equation}\label{eq:chi2eff}
    \chi^2_{\mu,eff}=\bar{\chi}^2_{\mu}-2\ln(B_\mu f_\mu)\,.
\end{equation}
We associate a weight to each knot $w_\mu\propto\exp(-\chi^2_{\mu,eff}/2)$. The knot at which this quantity is maximum or, equivalently, at which $\chi^2_{eff}$ is minimum, is associated to the best-fit values of ($M, \Omega_k$).

The two-dimensional probability for the parameters $X$ and $Y$, $P_{XY}$, can be easily computed as follows,

\begin{equation}
P_{XY}(x,y)=\frac{w_{\mu\to (x,y) }}{\sum\limits_{\beta} w_\beta}
\end{equation}
where in the denominator we sum over all the knots, and in the numerator we only consider the knot associated to the values $x$ and $y$ of the parameters $X$ and $Y$, respectively. 

We can also compute the one-dimensional posterior probability for each parameter $X$, $P_X$, using the analogous expression

\begin{equation}\label{eq:P_chi2eff}
    P_X(x)=\frac{\sum\limits_{\mu\to x} w_\mu}{\sum\limits_{\beta} w_\beta}\,,
\end{equation}
where now in the numerator we sum over those knots associated to the value $x$ of the parameter $X$.

The one-dimensional posteriors and the confidence regions at 68\% and 95\% C.L. in all the planes of parameter space are provided in Fig. \ref{fig:joint3D}. By evaluating for each parameter the maximum of the probability Eq. \eqref{eq:P_chi2eff} and the 68\% confidence intervals, we obtain the following results: $M=(-19.344^{+0.116}_{-0.090})$ mag and $\Omega_k=-0.07^{+0.27}_{-0.21}$. The constraint on $\Omega_k$ is similar to the one found by \cite{Dhawan:2021mel} using the Pantheon compilation of SNIa (instead of the most updated Pantheon+) and without considering the correlations between the CCH data nor the data point from \citep{Borghi:2021rft}, $\Omega_k=-0.03\pm 0.26$. In addition, we also provide a constraint on $M$, which is not reported by \cite{Dhawan:2021mel}, since they marginalize their result over it.

\begin{figure*}
    \centering
    \includegraphics[width=5.5in, height=5.5in]{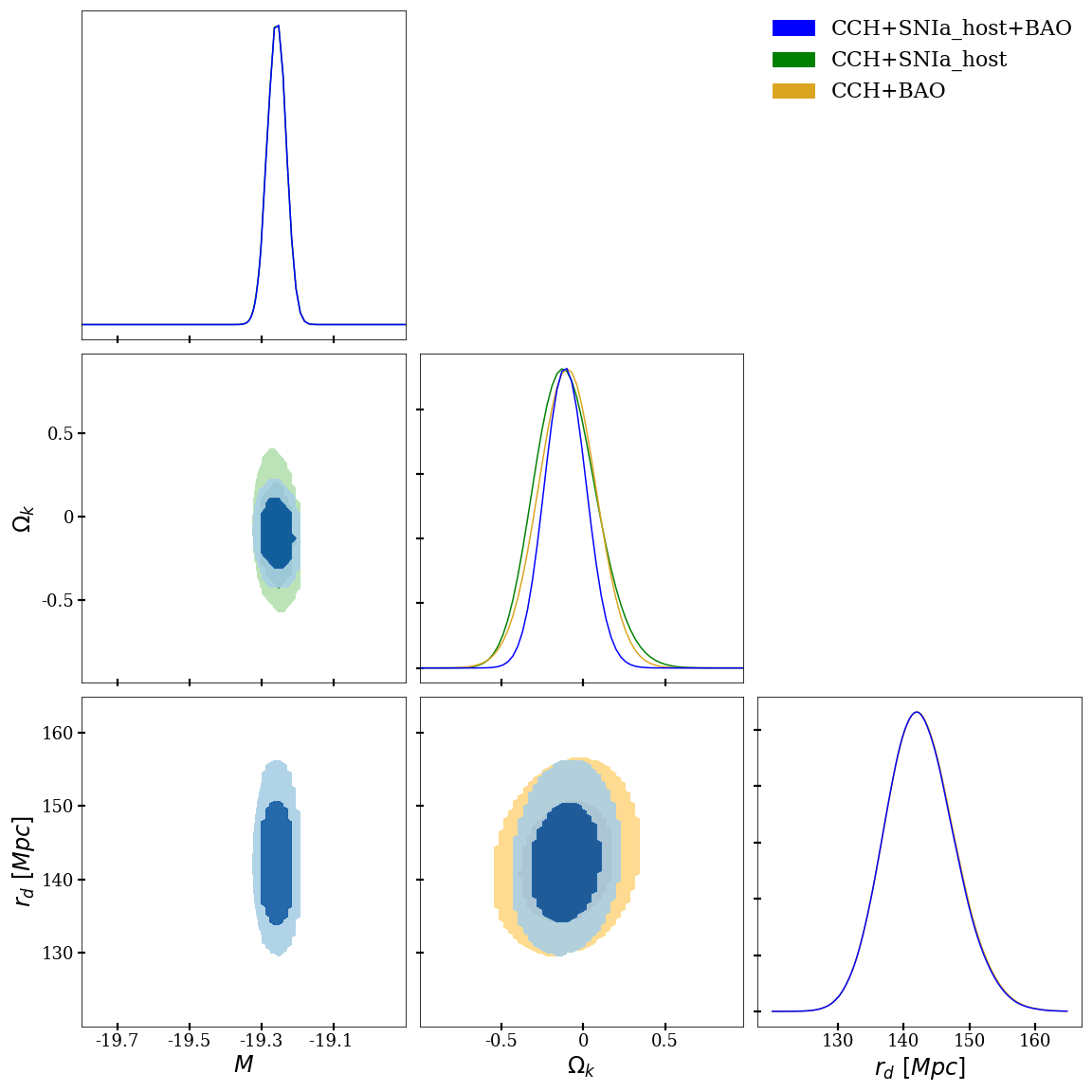}
    \caption{As in Fig. \ref{fig:joint3D}, but including the information from the SNIa in the host galaxies \citep{Scolnic:2021amr} and their distances employed by SH0ES \citep{Riess:2021jrx}. See Sec. \ref{sec:SNIahost} for further comments.}
    \label{fig:joint3DSH0ES}
\end{figure*}


\subsection{Analysis with CCH+BAO}\label{sec:CCH+BAO}
The same methodology described in Sec. \ref{sec:CCH+SN} can be applied in an analogous way to the plane $(\Omega_k,r_d)$ using the CCH and the BAO data sets. The former is used to calculate $N$ vectors with the values of $H(z)$ and the angular diameter distance $D_A(z)$ at the BAO redshift points. This information can then be employed to perform a $\chi^2$ analysis and compute the weights at each point of the grid. In this case the grid ranges are $\Omega_k \in [-1,1]$ and $r_d \in [120, 165]$ Mpc.

The resulting constraints read $\Omega_k=-0.10\pm0.18$ and $r_d=(141.9^{+5.6}_{-4.9})$ Mpc. The combination of the CCH data set with the BAO measurements still favors a negative central value for the curvature parameter, although it is compatible with a flat geometry within only $\sim0.6\sigma$. The uncertainty of $\Omega_k$ is $\sim25\%$ smaller than in the analysis with CCH+SNIa. This is clear from the comparison of the green and yellow one-dimensional posteriors of the curvature parameter in Fig. \ref{fig:joint3D}.

The posterior of $r_d$ peaks $\sim 1\sigma$ below the preferred {\it Planck}/$\Lambda$CDM value, $r_d=(147.09\pm 0.26)$ Mpc, closer to the region preferred by Early Dark Energy and modified gravity models proposed to alleviate the $H_0$ tension, see e.g. \citep{Poulin:2018cxd,SolaPeracaula:2020vpg}. However, this deviation is not statistically significant.


\subsection{Joint analysis with CCH+SNIa+BAO}\label{sec:CCH+SN+BAO}
The full parameter space can be now explored to get joint constraints for $(M,\Omega_k,r_d)$ by taking advantage of the results gathered in Secs. \ref{sec:CCH+SN} and \ref{sec:CCH+BAO}. We can combine the previous results to get a total $\chi^{2}$ as follows,

\begin{equation}\label{eq:chi2_3D}
    \chi^{2}(M,\Omega_k, r_d) = \chi^{2}(M,\Omega_k)+\chi^{2}(\Omega_k, r_d)\,,
\end{equation}
since the SNIa and BAO data are independent. We use again the expression \eqref{eq:P_chi2eff}  to obtain the individual constraints for the three parameters. The final results read: $M=(-19.314^{+0.086}_{-0.108})$ mag, $\Omega_k=-0.07^{+0.12}_{-0.15}$ and $r_d=(142.3\pm5.3)$ Mpc. As expected, the combination of all the low-$z$ data sets employed in this work leads to smaller uncertainties (see Table \ref{tab:joint} and Fig. \ref{fig:joint3D}), specially in the case of $\Omega_k$, since this is the only parameter that is constrained from both the CCH+SNIa and CCH+BAO data sets. If we set $\Omega_k=0$ we find $M=(-19.346^{+0.094}_{-0.088})$ mag and $r_d=(142.6\pm5.3)$ Mpc, which remain extremely close to the main results, but with slightly smaller errors\footnote{We take the arithmetic mean of the upper and lower uncertainties to make this comparison.}.

Our result for the absolute magnitude is independent of the SNIa distance ladder calibration. It is compatible within 1$\sigma$ with $M^{R22}$, but our method cannot achieve the precision attained by SH0ES \citep{Riess:2021jrx}. We study in Sec. \ref{sec:SNIahost} the impact of considering also the SNIa in the host galaxies and their distances.
Our value of $M$ is also in agreement with the one in \citep{Gomez-Valent:2021hda}, obtained using a different method based on the index of inconsistency by \cite{Lin:2017ikq}, the Pantheon data set and less CCH data points, but still making use of the combination CCH+BAO+SNIa. We find a 1$\sigma$-compatibility also with the $\Lambda$CDM result $M=(-19.403\pm 0.010)$ mag \citep{Gomez-Valent:2022hkb}, although we remark again that our results have been obtained in a model-independent way.

Our measurement of $\Omega_k$ points very mildly to a closed universe, being compatible with the flatness assumption within only $\sim0.5\sigma$. In contrast to the previous work \citep{Gomez-Valent:2021hda}, which reports $\Omega_k = -0.01 \pm 0.1$, here we do not make use of any cosmological prior inspired by the {\it Planck}/$\Lambda$CDM results. The latter would dominate the final constraint on the curvature parameter over the low-$z$ data sets, something that we wanted to avoid here. The uncertainty of $\Omega_k$ is much larger than the one obtained in model-dependent analyses like the one by \cite{Aghanim:2018eyx} or \cite{Vagnozzi:2020dfn}. The latter obtain $\Omega_k = -0.0054 \pm 0.0055$ in the context of the non-flat $\Lambda$CDM by combining the CMB data from {\it Planck} with CCH.

The calibration of the standard ruler with CCH+SNIa+BAO leads to a value which is $\sim1\sigma$ smaller than the $\Lambda$CDM value preferred by {\it Planck} \citep{Aghanim:2018eyx}, similar to the one obtained from the  CCH+BAO analysis, and again peaks at values more in accordance with theoretical scenarios that alleviate the Hubble tension. Our value of the sound horizon at the drag epoch is also compatible with other model-independent analyses, as those by \cite{Haridasu:2018gqm}, $r_d=(145.6\pm 5)$ Mpc, and \cite{Gomez-Valent:2021hda}, $r_d=(146.0^{+4.2}_{-5.1})$ Mpc.

We also measure $H_0$ employing as a prior our CCH+SNIa+BAO constraint on $M$ and the apparent magnitudes of the SNIa
in the Hubble flow ($0.023<z<0.15$). We make use of the cosmographical expansion

\begin{equation}
D_L(z)=\frac{cz}{H_0}\left[1+\frac{z}{2}\left(1-q_0\right)\right]+\mathcal{O}(z^3)\,.
\end{equation}
Curvature corrections are of third order in $z$ and, hence, we can neglect them in this analysis. We obtain $H_0=(71.5\pm 3.1)$ km/s/Mpc, with an uncertainty that is roughly a factor $1/2$ smaller than the one obtained using only CCH, see Sec. \ref{sec:recHm}. As a byproduct, we also constrain the deceleration parameter $q_0=-0.44\pm 0.19$. This result is fully compatible with the model-independent measurements extracted from CCH+SNIa+BAO \citep{Haridasu:2018gqm,Gomez-Valent:2018gvm}, but with an uncertainty a factor two larger, since here $q_0$ is fixed only by the SNIa in the Hubble flow.

Our results are independent of the direct and inverse distance ladders, quite model-independent and robust under the use of alternative GP kernels (cf. Appendix \ref{sec:AppendixB}). This is interesting {\it per se}. However, they cannot arbitrate the $H_0$ tension yet. The low-redshift data sets under consideration give still room to new physics both in the pre- and post-recombination eras.


\begin{table}
    \centering
    \caption{The same as in Table \ref{tab:joint}, but including the SNIa in the host galaxies \citep{Scolnic:2021amr} and their distances to calibrate the SNIa as SH0ES \citep{Riess:2021jrx}. This has a very little impact on our constraints on $\Omega_k$ and $r_d$.}
    \label{tab:joint2}
    \begin{tabular}{cccc}        
         & {\scriptsize CCH+SNIa\_host} & {\scriptsize CCH+BAO} & {\scriptsize CCH+SNIa\_host+BAO} \\  
        \hline
        \hline
        M [mag]& $-19.252^{+0.024}_{-0.036}$ &  & $-19.252^{+0.024}_{-0.036}$  \\   
         \hline
         $\Omega_k$ & $-0.13^{+0.18}_{-0.21}$ & $-0.10\pm0.18$ & $-0.10^{+0.12}_{-0.15}$ \\
         \hline     
         $r_d$ [Mpc] & & $141.9^{+5.6}_{-4.9}$& $141.9^{+5.6}_{-4.9}$ \\      
        \hline
    \end{tabular}
\end{table}

\subsection{Considering smaller uncertainties in the CCH data}
\label{sec:spec}
The GPs kernel performance test done in Sec. \ref{sec:selectGP} shows that the mean values of the reduced chi-square, $\tilde \chi^{2}_{\mu}=\chi_\mu^2/{\rm dof}$ Eq. \eqref{eq:kernel_chi2}, associated to the reconstruction of $H(z)$ with the CCH data points listed in Table \ref{tab:CCH} are all much smaller than 1 (see also Fig. \ref{fig:histcov}). This result is not expected to be due to an overfitting of the GP, since similar values of the  $\tilde{\chi}^2$ are also found in fitting analyses with a simple straight line or a parabola, cf. Table 5 of \citep{Gomez-Valent:2018hwc}. As already mentioned, the small values of $\tilde{\chi}^2_\mu$ could instead be a hint of an overestimation of the errors in the covariance matrix of the CCH data, $\tilde C_{ij}$.
In this section, we want to explore this possibility by studying how the results in the analyses of Secs. \ref{sec:CCH+SN}-\ref{sec:CCH+SN+BAO} change if we allow for smaller uncertainties in $\tilde C_{ij}$. With this aim we build the new CCH covariance matrix $\tilde C_{ij,new} = \tilde C_{ij} / N^2$, with $N$ a positive factor. This is equivalent to decrease all the individual CCH uncertainties by a common factor $N$, while leaving the previous correlation coefficients intact. For this purpose, we first repeat the test of Sec. \ref{sec:selectGP} with the Matérn 32 kernel, but increasing the values of $N>1$ until the mean of the corresponding reduced chi-squared equals one, i.e. until $\tilde \chi^{2}_{\mu}=1$. We find that this happens when the CCH uncertainties decrease by a factor $N=1.579\sim 3/2$. We denote the resulting CCH data set with the new covariance matrix simply as CCH$_{\rm new}$ to distinguish it from the original one (CCH).
We can now repeat the analyses of Secs. \ref{sec:CCH+SN}-\ref{sec:CCH+SN+BAO} with CCH$_{\rm new}$ to study the impact of this change on the data uncertainties, bearing in mind that this is only a first (naive) attempt to estimate the impact of a possible overestimation of the uncertainties of the CCH data\footnote{A more refined analysis would probably require a better understanding of the systematics in the data and/or the application of an improved statistical method, on the lines of what was done by \cite{Hobson:2002zf}.}. This leads to the following CCH$_{\rm new}$+SNIa+BAO constraints: $M = (-19.326^{+0.050}_{-0.068})$ mag, $\Omega_k = 0.10^{+0.12}_{-0.15}$ and $ r_d = (142.6^{+3.9}_{-3.5})$ Mpc. The uncertainties of $r_d$ and $M$ decrease by a $\sim$30\%-40\% with respect to those found in the CCH+SNIa+BAO analysis (see also Fig. \ref{fig:joint3D}). We do not find, however, the same decrease in the uncertainty of $\Omega_k$. The reason is simple. Let us focus on the combination CCH$_{\rm new}$+SNIa. The low-redshift data at $z\ll1$ basically constraints $M$ and is insensitive to the curvature parameter. At larger redshifts, though, we can get constraints on $\Omega_k$, which depend on the reconstruction of the ratio $E(z)=H(z)/H_0$, see formula \eqref{eq:DL}. The point is that the correlation coefficients employed in the new CCH data set are exactly the same as those used in the original analysis, what makes the reconstructed shape of $E(z)$ to remain the same. This fact, in turn, explains why we find the same constraint on $\Omega_k$ as well.   

\begin{figure*}
    \centering
    \includegraphics[width=\columnwidth]{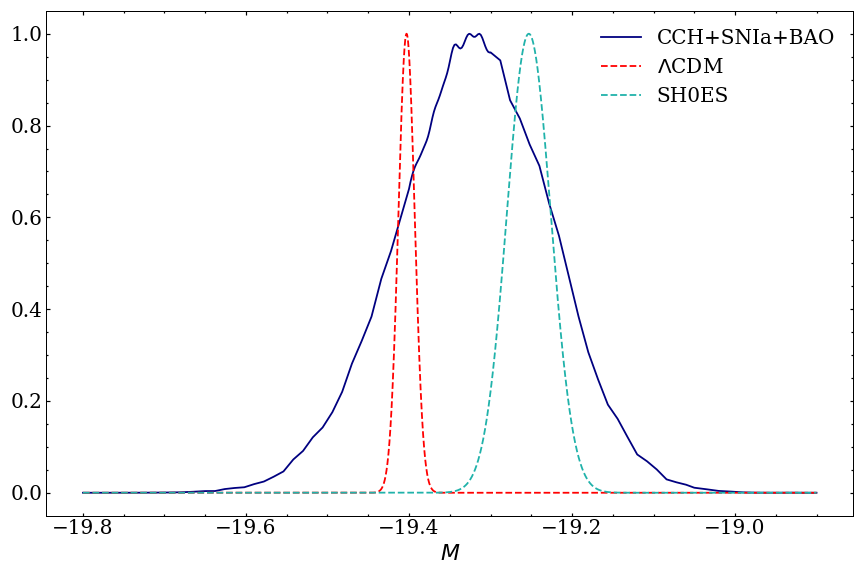}
    \includegraphics[width=\columnwidth]{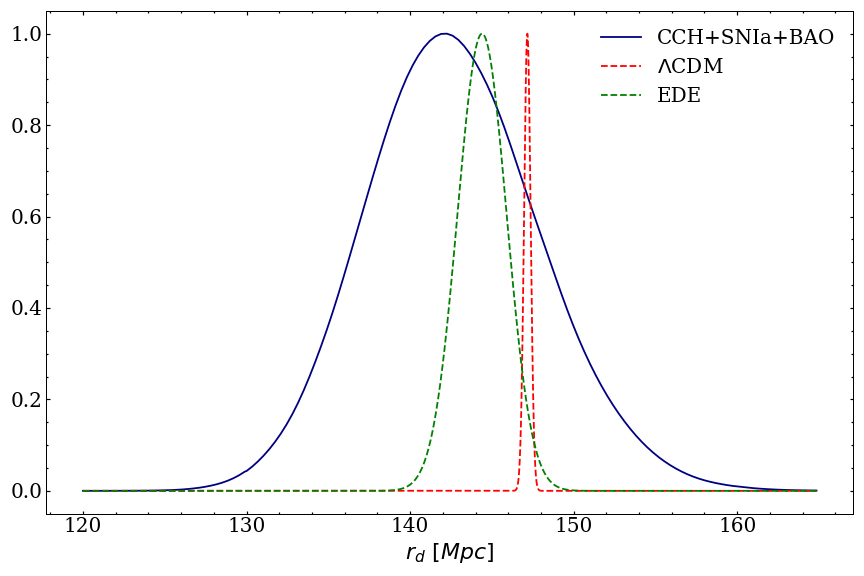}
    \includegraphics[width=\columnwidth]{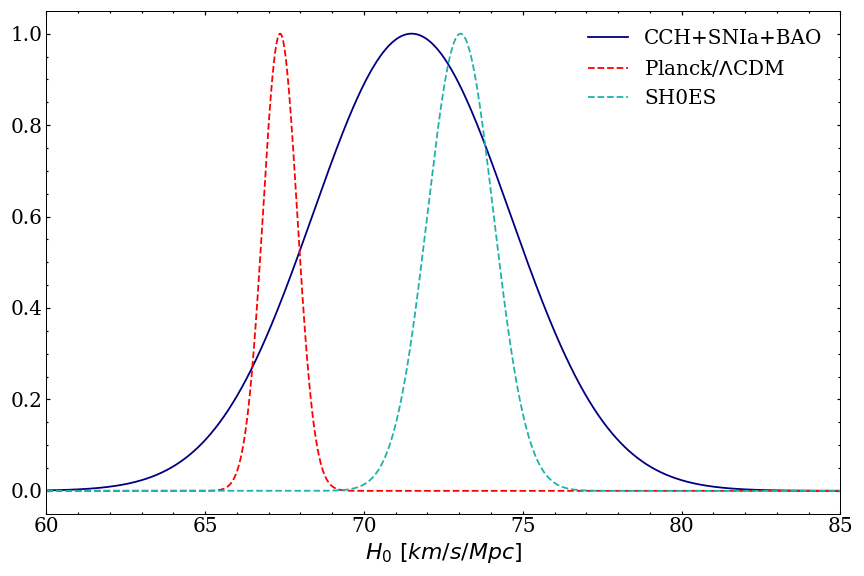}
    \caption{{\it Upper left plot:} Model-independent constraint on $M$ obtained from the analysis of Sec. \ref{sec:CCH+SN+BAO}, compared with the SH0ES posterior $M^{R22}=(-19.253\pm0.027)$ mag \citep{Riess:2021jrx} and the constraint obtained in the $\Lambda$CDM from the fitting analysis with Planck2018+SNIa+BAO data, $M=(-19.403\pm0.010)$ mag \citep{Gomez-Valent:2022hkb}. {\it Upper right plot:} The same, but for $r_d$. Here, our result is compared with the posterior obtained from the Planck2018+SNIa+BAO analyses in the context of $\Lambda$CDM, $r_d=(147.17\pm0.20)$ Mpc, and the ultra-light axion-like model of early dark energy, $r_d=(144.4\pm1.5)$ Mpc \citep{Gomez-Valent:2022hkb}. {\it Lower plot:} Constraint on $H_0$ obtained from our CCH+SNIa+BAO prior on $M$ (cf. the upper left plot and Table \ref{tab:joint}) and the apparent magnitudes of the SNIa in the Hubble flow, cf. Sec. \ref{sec:CCH+SN+BAO}. This result is compared with the SH0ES \citep{Riess:2021jrx} and {\it Planck}/$\Lambda$CDM \citep{Aghanim:2018eyx} values.}
    \label{fig:results_comp}
\end{figure*}


\subsection{Inclusion of the SNIa in the host galaxies and their distances}\label{sec:SNIahost}

In our main analyses of Secs. \ref{sec:CCH+SN}-\ref{sec:CCH+SN+BAO}, and also in Sec. \ref{sec:spec}, we have excluded the SNIa located in the Cepheid host galaxies, i.e. those employed by SH0ES to calibrate the SNIa in the second rung of the cosmic distance ladder \citep{Riess:2021jrx, Scolnic:2021amr}. We do so to obtain results that are independent of the main drivers of the Hubble tension. Nevertheless, we may ask ourselves which is the impact of considering this additional information, which actually is included in the full Pantheon+ compilation. We call this SNIa data set SNIa\_host, in short, and follow the same procedure applied in Secs. \ref{sec:CCH+SN}-\ref{sec:CCH+SN+BAO}. The results of this analysis are shown in Fig. \ref{fig:joint3DSH0ES} and listed in Table \ref{tab:joint2}. The output from the analysis with CCH+BAO is not sensitive to the changes in the SNIa data set, for obvious reasons. As expected, the constraints on $M$ are fully dominated by the calibration of the SNIa at the host galaxies. In particular, for the CCH+SNIa\_host+BAO analysis we obtain: $M=(-19.252^{+0.024}_{-0.036})$ mag, $\Omega_k=-0.10^{+0.12}_{-0.15}$ and $r_d=(141.9^{+5.6}_{-4.9})$ Mpc, with $H_0=(74.0^{+1.0}_{-0.9})$ km/s/Mpc. No important differences are found in the curvature parameter and $r_d$ with respect to the results presented in Sec. \ref{sec:CCH+SN+BAO}.


\section{Conclusions}\label{sec:conclusions}

In this paper we have first reconstructed the absolute magnitude of SNIa and the curvature of the universe as a function of the redshift up to $z\approx2$ making use of Gaussian Processes and data on cosmic chronometers and the Pantheon+ compilation of supernovae of Type Ia. We have found that these low-redshift data sets do not point to a time evolution of the SNIa intrinsic luminosity nor a breaking of the homogeneity of the universe at large scales. Both, $M(z)$ and $\Omega_k(z)$ are compatible at $68\%$ C.L. with a constant. In addition, we have also tested the consistency of the BAO data from the galaxy surveys 6dFGS, BOSS, eBOSS, WiggleZ and DES Y3, by checking that they are all compatible with a common value of $r_d$, at least under the precision offered by the CCH data. Motivated by these results, we have then constrained with CCH, SNIa and BAO the constant values of $\Omega_k$ and the calibrators of the direct and inverse distance ladders, $M$ and $r_d$. We have done so by applying a quite model-independent method, which is also independent from the first rungs of the cosmic distance ladder employed by SH0ES and the CMB data from {\it Planck}, i.e. from the main data sets involved in the Hubble tension. This is in contrast to other results obtained in the context of the $\Lambda$CDM, see e.g. \citep{Aghanim:2018eyx,Handley:2019tkm,DiValentino:2019qzk,Gomez-Valent:2022hkb}. We obtain: $\Omega_k=-0.07^{+0.12}_{-0.15}$, $M=(-19.314^{+0.086}_{-0.108})$ mag and $r_d=(142.3\pm 5.3)$ Mpc. We have checked that the inclusion of the SNIa in the host galaxies and their distances only affects the value of $M$, making its central value and uncertainties to be very close to those measured by SH0ES. 

 \begin{figure*}
    \centering    \includegraphics[width=6.5in, height=3in]{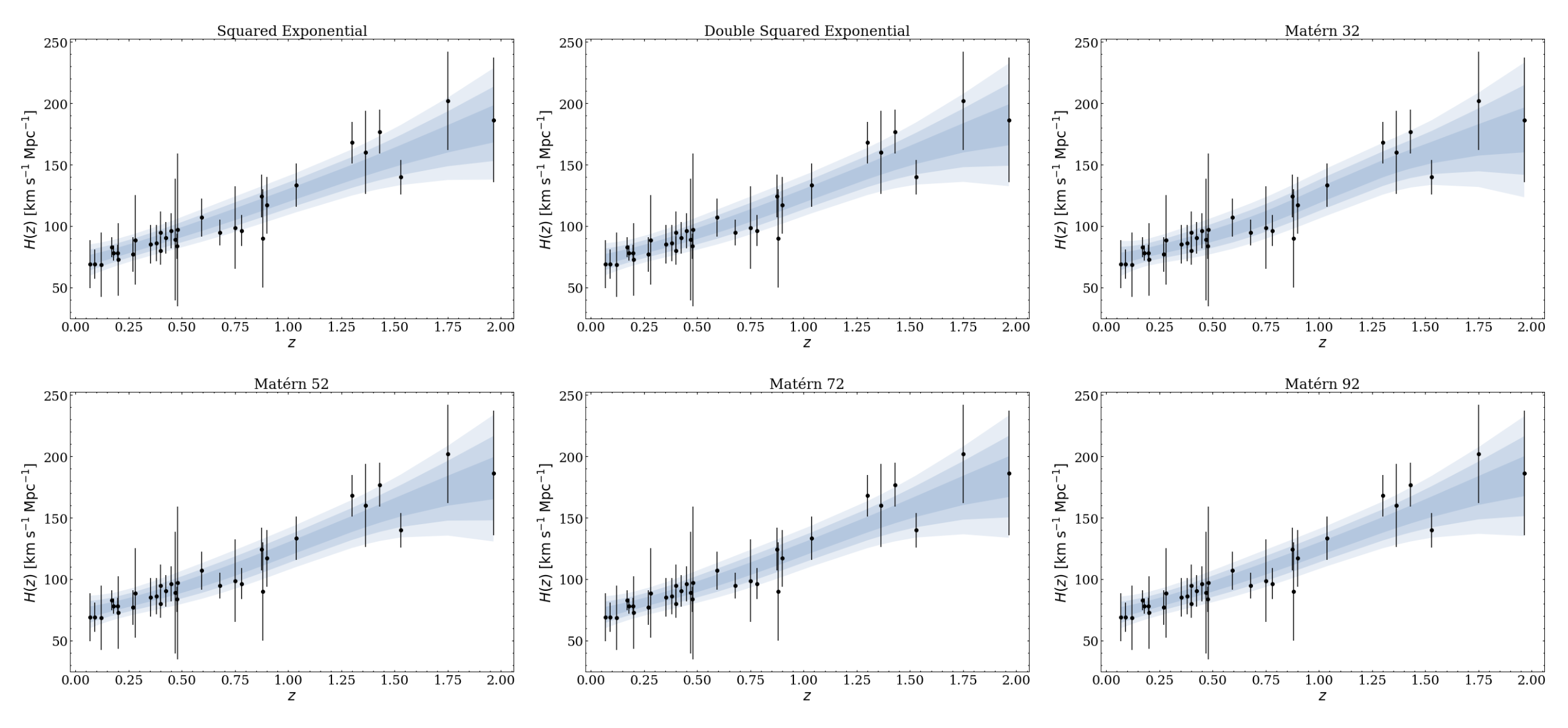}
    \caption{Reconstructed shapes of the Hubble function $H(z)$ at 1$\sigma$, 2$\sigma$ and 3$\sigma$ C.L. obtained from Gaussian Processes and the CCH data of Table \ref{tab:CCH} (in black) with the six different GP kernels described in Sec. \ref{sec:selectGP}.}
    \label{fig:kernelsperf}
\end{figure*}

Our results improve previous constraints in the literature obtained also with Gaussian Processes but with slightly different data sets and methodologies. For instance, \cite{Benisty:2022psx} obtained $M=(-19.42\pm 0.35)$ mag from data on BAO and SNIa together with a {\it Planck} prior for $r_d$. We, instead, have measured $M$ with an uncertainty three times smaller. In addition, we have extracted joint and model-independent constraints for $\Omega_k$ and $r_d$ as well, with a more refined BAO data set, which is free from double-counting issues and incorporates the effect of correlations. Our determination of $\Omega_k$ is $50\%$ more precise than the one carried out by \cite{Dhawan:2021mel}, $\Omega_k=-0.03\pm 0.26$, thanks mainly to the use of BAO data on top of the CCH and the Pantheon+ compilation of SNIa. The same level of improvement is also obtained compared to the cosmographical analysis of SNIa and strong lensing data by \cite{Collett:2019hrr}, who reported $\Omega_k=0.12^{+0.27}_{-0.25}$. The present work also improves the analysis of \citep{Gomez-Valent:2021hda}, since here we have not used any external prior for the curvature parameter and have employed the SNIa contained in the Pantheon+ compilation, instead of those of Pantheon. However, the uncertainties that we have found are still one order of magnitude larger compared to the model-dependent determinations by {\it Planck} \citep{Aghanim:2018eyx}. As discussed by \cite{Dhawan:2021mel}, this could change in the next years, when e.g. SNIa data from the Vera C. Rubin Observatory's Legacy Survey of Space and Time (LSST, \citealt{LSSTScience:2009jmu,LSST:2008ijt}) and BAO data from Euclid \citep{EUCLID:2011zbd} and the Dark Energy Spectroscopic Instrument (DESI, \citealt{DESI:2016fyo}) become available. This will not only decrease the uncertainties of the curvature parameter through the model-independent analyses of standard candles and large-scale structure data \citep{Amendola:2019lvy,Amendola:2022vte}, but also improve the constraints we get for the calibrators $M$ and $r_d$, which is obviously important for the discussion of the $H_0$ tension. As shown in Fig. \ref{fig:results_comp}, in the light of the current low-redshift data our method does not let us arbitrate the tension yet (we obtain $H_0=(71.5\pm 3.1)$ km/s/Mpc with CCH+SNIa+BAO), but we might be able to do so with the advent of the aforementioned upcoming telescopes and surveys. We have seen in Sec. \ref{sec:spec} that a decrease by a factor 3/2 of the uncertainties of the CCH data 
produces a $30\%-40\%$ decrease of the uncertainties of the calibrators. Thus, an improvement in the CCH data, either in terms of quality or quantity, can also have a non-negligible impact on the outcome of this method. Euclid, for instance, is expected to provide up to a few thousands passively evolving galaxies at $z\lesssim 2$, increasing by 2 orders of magnitude the currently available statistics \citep{Moresco:2022phi}.

The method we propose will then find interesting applications in the future, when all these new data become a reality. It will serve as both a discriminator of models beyond the $\Lambda$CDM and an independent means of testing the calibration of the direct and inverse cosmic distance ladders.

\appendix

\section{GP-reconstruction of \texorpdfstring{$H(z)$}{} with different kernels}\label{sec:AppendixA}

In Sec. \ref{sec:selectGP} we have explained a method to select in an objective way a GP-kernel among a group of them given a collection of data points. Here, we just show in Fig. \ref{fig:kernelsperf} the reconstructed shapes of the Hubble function obtained from six different kernels, namely: Squared Exponential, Double Squared Exponential, Matérn 32, Matérn 52, Matérn 72 and Matérn 92. As already discussed, the differences are not important. This resonates well with the results reported in Table \ref{tab:kernel_perform} and the conclusions reached in Sec. \ref{sec:selectGP}.


\section{Results with the Gaussian kernel}\label{sec:AppendixB}

In this appendix we briefly study the robustness of the results presented in Secs. \ref{sec:CCH+SN}-\ref{sec:CCH+SN+BAO} under the choice of a different GP kernel. To do so, we adopt the Gaussian kernel, which is defined by Eq. \eqref{eq:SE}. It is the smoothest kernel within the Matérn family. It is infinitely differentiable, and so is also the reconstructed function obtained from the GP, see Sec. \ref{sec:selectGP} for details. Using the Gaussian kernel instead of Matérn 32 we find the following results with the compilation of data CCH+SNIa+BAO: $M=-19.314^{+0.098}_{-0.079}$ mag, $\Omega_k=-0.10^{+0.12}_{-0.15}$ and $r_d=141.9^{+5.3}_{-4.9}$ Mpc. By comparing these results to those provided in Table \ref{tab:joint} we see that they are stable under the choice of the kernel. The shift in the central value of $\Omega_k$ is equal to the bin size of the grid, whereas we find a $4\%$ decrease in the error bars of $r_d$ and a $9\%$ in $M$, and completely compatible results also for the central values of these parameters. In our main analysis we opt, though, to employ Matérn 32, since this is the kernel that leads to the most conservative results, cf. again Sec. \ref{sec:selectGP}.


\section*{Acknowledgments} 
AGV is funded by the Istituto Nazionale di Fisica Nucleare (INFN) through the project of the InDark INFN Special Initiative: ``Dark Energy and Modified Gravity Models in the light of Low-Redshift Observations'' (n. 22425/2020).  He also acknowledges the participation in the COST Action CA21136 “Addressing observational tensions in cosmology
with systematics and fundamental physics” (CosmoVerse). The authors acknowledge support by the INFN project “InDark”. MM is also supported by the ASI/LiteBIRD grant n. 2020-9-HH.0 and by the  Fondazione  ICSC, Spoke 3 Astrophysics and Cosmos Observations, National Recovery and Resilience Plan (Piano Nazionale di Ripresa e Resilienza, PNRR) Project ID CN\_00000013 "Italian Research Center on High-Performance Computing, Big Data and Quantum Computing"  funded by MUR Missione 4 Componente 2 Investimento 1.4: Potenziamento strutture di ricerca e creazione di "campioni nazionali di R\&S (M4C2-19 )" - Next Generation EU (NGEU).


\section*{Data availability}

The data employed in this article are publicly available (see Sec. \ref{sec:data} and references therein) and our codes will be shared on reasonable request.


\bibliographystyle{mnras}
\bibliography{Calib_Ok.bib}

\end{document}